\documentclass[10pt]{article}

\usepackage{amsmath}
\usepackage{amssymb}

\usepackage{graphicx}

\usepackage{cite}

\usepackage{color} 

\usepackage[labelfont=bf,labelsep=period,justification=raggedright]{caption}

\makeatletter
\renewcommand{\@biblabel}[1]{\quad#1.}
\makeatother

\date{}

\newcommand{\AppendFigures}[1]{}	

\newcommand{\script}[1]{{\mbox{\scriptsize #1}}}

\newcommand{\CITE}[1]{ \cite{#1}}
\newcommand{\VEC}[1]{\mathbf{#1}}
\newcommand{\Eq}[1]{Eq. {#1}}
\newcommand{\Eqs}[1]{Eqs. {#1}}

\newcommand{\EqPunc}[1]{}
\newcommand{\EqPeriod}[1]{}
\newcommand{\Fig}[1]{Fig. {#1}}

\newcommand{\Figs}[1]{Figs. {#1}}
\newcommand{\SFig}[1]{Fig. {#1}}

\newcommand{\Table}[1]{Table {#1}}

\newcommand{\FigureInText}[1]{}

\newcommand{\FigureInLegends}[1]{}
\newcommand{\TableInLegends}[1]{#1}
\renewcommand{\VEC}[1]{\mbox{\boldmath{$#1$}}} 

\renewcommand{\FigureInText}[1]{}

\renewcommand{\FigureInLegends}[1]{#1}
\renewcommand{\FigureInLegends}[1]{}
\renewcommand{\TableInLegends}[1]{#1}

\newcommand{\FULLorSHORT}[2]{{#2}}       

\AppendFigures{
\topmargin -1.0cm
\pagestyle{empty}	
} 

\newcommand{\BF}[1]{\textbf{#1}}

\renewcommand{\SFig}[1]{Fig. {#1}}

\newcommand{\TextFig}[1]{#1}
\newcommand{\SupFig}[1]{}
\newcommand{\TextTable}[1]{#1}
\newcommand{\SupTable}[1]{}

\begin{document}


\noindent
\begin{flushleft}
{\Large
\textbf{
%
%
%
Inference of Co-Evolving Site Pairs:
an Excellent Predictor of Contact Residue Pairs in Protein 3D structures
}
}
\\
\vspace*{1em}
Sanzo Miyazawa
\\
Graduate School of Engineering, Gunma University, Kiryu, Gunma 376-8515, Japan
\\
E-mail: sanzo.miyazawa@gmail.com
\end{flushleft}

\section*{Abstract}
Residue-residue interactions that fold a protein into a unique three-dimensional structure
and make it play a specific function          
impose structural and functional constraints on each residue site.
Selective constraints on residue sites are recorded 
in amino acid orders in homologous sequences 
and also in the evolutionary trace of amino acid substitutions. 
A challenge is to extract direct
dependences between residue sites 
by removing indirect dependences
through other residues within a protein or even through other molecules.
Recent attempts of disentangling direct from indirect dependences of amino acid types 
between residue positions in multiple sequence alignments have revealed that
the strength of inferred residue pair couplings is
an excellent predictor of residue-residue proximity in folded structures.
Here, we report an alternative attempt of inferring 
co-evolving site pairs
from concurrent and compensatory substitutions between sites 
in each branch of a phylogenetic tree.
First, branch lengths of a phylogenetic tree inferred 
by the neighbor-joining method are optimized as well as other parameters 
by maximizing a likelihood of the tree in a 
mechanistic codon substitution model.
Mean changes of quantities, which are characteristic of
concurrent and compensatory substitutions, accompanied by substitutions
at each site in each branch of the tree are estimated 
with the likelihood of each substitution.
Partial correlation coefficients of 
the characteristic changes 
along branches 
between sites,
in which linear dependences on other sites are removed,
are calculated and used to rank co-evolving site pairs.
Accuracy of contact prediction 
based on the present co-evolution score
is comparable to that achieved by a maximum entropy model of protein sequences 
for 15 protein families taken from the Pfam release 26.0.
Besides, this excellent accuracy indicates that
compensatory substitutions are significant in protein evolution.



\section*{Introduction}

The evolutionary history of protein sequences
is a valuable source of information in many fields of science
not only in evolutionary biology but even to understand protein structures.
Residue-residue interactions that fold a protein into a unique three-dimensional (3D) structure
and make it play a specific function 
impose structural and functional constraints on each amino acid.
Selective constrains on amino acids are recorded
in amino acid orders in homologous protein sequences
and also in the evolutionary trace of amino acid substitutions.
Negative effects caused by mutations at one site 
must be compensated by successive mutations at other sites\CITE{YHT:64,FM:70,BKOSK:04,MA:04},
otherwise negative mutants 
will be eliminated from a gene pool and never reach fixation in a population.
Such structural and functional constraints arise from interactions between sites 
mostly in close spatial proximity. 
Thus, it is suggested that the types of amino acids  
and amino acid substitutions 
must be correlated between sites that are close in a protein 3D structure
\CITE{AVBMN:88,GSSV:94,SKS:94,PT:97,PTG:99,AWFTD:00,FOVC:01,FA:04,MGDW:05,FT:06,GP:07,DWG:08,SPSLAGL:08,WWSHH:09}.
However, correlations of amino acid types and amino acid substitutions do not only reflect
direct interactions between sites but also indirect ones through other residues
within a protein or even through other molecules involved in
a molecular complex\CITE{LR:99,YH:07} such as oligomerization\CITE{DWG:08}, 
protein-substrate, protein-protein\CITE{WWSHH:09}, and protein-DNA.
In addition, statistical noise due to the small number of samples and
methodological limitations are obstacles to decode correlations into spatial
relationships between sites.
However, protein families consisting of homologous sequences 
in a wide range of divergence
are now collected in protein family databases such as Pfam \CITE{PCEMTBPFCCHHSEBF:12}, 
and become available to reduce statistical noise to a sufficiently small amount.
A present challenge is thus to extract only direct dependences between sites
by excluding indirect correlations between them
from a wide variety of homologous sequences evolutionarily exploited in a sequence space
\CITE{PLFP:08,BN:08,BN:10,MPBMSZOHW:11,MCSHPZS:11,TS:11}.

Extracting essential information from the evolutionary sequence record have been attempted
using global statistical models.
A Bayesian graphical model was applied to 
disentangling direct from indirect dependencies between residue positions
in multiple sequence alignments of proteins \CITE{BN:08},
and a significant improvement was achieved in the accuracy of contact prediction\CITE{BN:10}.
A Bayesian graphical model was also applied to
the analysis of the joint distribution of substitution events 
to identify significant associations among residue sites\CITE{PLFP:08}.
Recently, 
remarkable accuracy of contact prediction
was achieved\CITE{MPBMSZOHW:11,MCSHPZS:11} by 
using a maximum entropy model\CITE{WWSHH:09} of a protein sequence, 
constrained by the statistics of a multiple sequence alignment,
to infer residue pair coupling.
Partial correlation coefficients 
derived from mutual information of residue pair coupling 
were also used to extract direct information\CITE{TS:11}.
They developed not only a robust method to extract essential correlations of amino acid type
between residue positions in multiple sequence alignments, but
also showed that inferred residue-residue proximities provide sufficient information to
predict a protein fold without the use of known three-dimensional structures.

Here, we report an alternative approach of inferring 
co-evolving site pairs
from concurrent and compensatory substitutions between sites 
in each branch of a phylogenetic tree.
First, branch lengths of a phylogenetic tree inferred 
by the neighbor-joining (NJ) method\CITE{SN:87} are optimized
by maximizing a likelihood of the tree 
in a mechanistic codon substitution model \CITE{M:11a,M:11b}.  
The variation of selective constraints over sites is approximated by
a discrete gamma distribution \CITE{Y:94}.
Mean changes of the various types of quantities,
which are characteristic of concurrent and compensatory substitutions,
accompanied by substitutions at each site in each branch of the tree 
are estimated on the basis of likelihoods. 
Partial correlation coefficients 
of their characteristic changes accompanied by substitutions
along branches
between sites are employed to remove
a linear dependence on characteristic changes along branches at other sites.
In other words, a Gaussian graphical model \CITE{E:00} is assumed 
for site dependence, 
because a conditional independence between two variables 
given other variables 
in a multi-variate Gaussian distribution is equivalent to
zero partial correlation coefficient between the two variables.
Then, co-evolution scores are defined from partial correlation coefficients
of the various types of the characteristic quantities.
Co-evolving site pairs are inferred in the decreasing order of 
the overall co-evolution score.
Accuracy of contact prediction based on the 
partial correlation coefficients
is comparable to that by a maximum entropy model\CITE{MCSHPZS:11} 
for 15 protein families taken from the Pfam release 26.0 \CITE{PCEMTBPFCCHHSEBF:12},
indicating that the present method can be an alternative approach for contact prediction.
Also, a fact that contact site pairs can be well predicted by the present method
strongly indicates that compensatory substitutions are significant in protein evolution,
because the present method based on concurrent and compensatory substitutions
will not work at all if all substitutions are completely neutral.


%


\section*{Results}

\noindent
\subsection*{Framework of the present method}
\vspace*{1em}

The framework of the present method is shown in \Fig{\ref{fig: framework}}.
For each protein family, its phylogenetic tree $T$ inferred by the neighbor-joining (NJ) method is
taken from the Pfam database\CITE{PCEMTBPFCCHHSEBF:12} and branch lengths $t_b$ of the tree are
optimized by maximizing the likelihood of the tree in a mechanistic codon substitution model.
Then, the average changes ($\Delta_{ib}$) of quantities, which are characteristic of
concurrent and compensatory substitutions, accompanied by substitutions at each site $i$
in each branch $b$ of the phylogenetic tree $\hat{T}$ are estimated
with the likelihood of each substitution.
Their correlation coefficients ($r_{\Delta_i \Delta_j}$) along branches between sites are
calculated, and converted to partial correlation coefficients, which
are correlation coefficients between residual vectors
($\Pi_{\bot\{\Delta_{k\neq i, j}\}} \VEC{\Delta}_i$ and $\Pi_{\bot\{\Delta_{k\neq i, j}\}} \VEC{\Delta}_j$)
of given two vectors that are perpendicular to a subspace
consisting of other vectors except those two vectors ($\VEC{\Delta}_i$ and $\VEC{\Delta}_j$)
and therefore cannot be accounted for by a linear regression on other vectors.
Finally, co-evolution scores based on the partial correlation coefficients 
are defined
and used to rank site pairs for close spatial proximity.

The following characteristic 
changes
defined in the Methods section 
are examined to detect concurrent and compensatory substitutions between sites;
(1) occurrence of amino acid substitution: $\VEC{\Delta}^s_i$, 
(2) side-chain volume: $\VEC{\Delta}^v_i$,
(3) side-chain charge: $\VEC{\Delta}^{c}_i$,
(4) hydrogen-bonding capability: $\VEC{\Delta}^{hb}_i$,
(5) hydrophobicity: $\VEC{\Delta}^{h}_i$,
(6) $\alpha$ propensity: $\VEC{\Delta}^{\alpha}_i$, 
(7) $\beta$ propensity: $\VEC{\Delta}^{\beta}_i$, 
(8) turn propensity: $\VEC{\Delta}^{t}_i$,
(9) aromatic interaction: $\VEC{\Delta}^{ar}_i$,
(10) branched side-chain: $\VEC{\Delta}^{br}_i$,
(11) cross-link capability: $\VEC{\Delta}^{cl}_i$,
and
(12) ionic side-chain: $\VEC{\Delta}^{ion}_i$.
All except the $\alpha$ propensity are used to define co-evolution scores.

\noindent
\subsection*{Protein families and sequences used}
\vspace*{1em}

In order to calculate partial correlation coefficients between sites by 
taking the inverse of
a covariance or correlation matrix, it must be regular so that the 
number of homologous sequences must be at least more than the number of sites, 
that is, the sequence length.  To obtain statistically reliable numbers, 
even more sequences than 10 times as many as sites would be needed.
In the Pfam database \CITE{PCEMTBPFCCHHSEBF:12}, 
protein domain families consisting of 
many thousands of homologous sequences are included.
Particularly, protein domain families used in \CITE{MCSHPZS:11}
to infer residue pair couplings
in multiple sequence alignments are appropriate to allow us to compare 
prediction accuracies between
the present method and their method.
These protein domain families 
in the Pfam release 26.0 (November 2011)
are listed in \Table{\ref{tbl: protein_families}}.
Also, \Table{\ref{tbl: protein_families}} shows 
the Uniprot ID and corresponding protein coordinates (PDB ID) 
of a target protein in each protein family, for which
co-evolving site pairs are predicted.

In the Pfam database,
there are two sets of  sequence alignments for each protein family;
a seed alignment and a full alignment.
Also, a phylogenetic tree inferred from each alignment
by the 
NJ
method \CITE{SN:87} are available.
Here the seed alignment and its phylogenetic tree 
are used to estimate parameters in
a mechanistic codon substitution model\CITE{M:11a,M:11b}.
With those parameters optimized for the seed alignment in the codon-based model,
posterior means of characteristic variables at each site 
in each branch of a phylogenetic tree
are estimated for subsets of a full alignment, after branch lengths are optimized.

The full alignments include closely-related sequences whose differences are
less than 0.01. The number of branches ($n_b$) in a phylogenetic tree is proportional to
the number of OTUs ($n_{\script{otu}}$) 
(operational taxonomic units that correspond to sequences in the present case); 
$n_b = 2 n_{\script{otu}} - 3$
for an unrooted tree. 
Computational time required for 
the present calculation increases with increasing number of branches. 
Including closely-related sequences
requires computationally intensive calculation, although
it is not much informative;
invariant sites do not have any information in the present method, which
is designed to detect concurrent and compensatory substitutions between sites in proteins.  
Thus, subsets made by excluding closely-related sequences from
the Pfam full alignments are used in the present calculations.
The subsets of a full alignment and their NJ trees
are made by removing OTUs that
are connected to the parent nodes with branches 
shorter 
than a certain threshold 
($T_{bt}$), 
although seed sequences and a target protein are not removed.

In addition, to reduce a computational load in the calculation of the likelihood of a 
phylogenetic tree, only site positions 
where amino acids are found in the target protein
are extracted from the multiple sequence alignment 
and used in the present analysis.

\noindent
\subsection*{Correlation versus partial correlation coefficients}
\vspace*{1em}

First, we examined how differently correlation coefficients and 
partial correlation coefficients between sites identify dependent site pairs.
The distribution of Pearson's correlation coefficient 
in the case of no correlation can be well approximated 
by the Student's $t$ distribution.  Therefore, here 
a correlation coefficient $r_t$ corresponding to
the E-value $E_t = 0.001$ (the P-value $P_t = E_t / n_{\script{pairs}}$) in the Student's t-distribution
of the degree of freedom $\mbox{df} = n_b - 2$ is used as
a threshold for significance; where $n_{\script{pairs}}$ is the number of site pairs and 
$n_b = 2 n_{\script{otu}} - 3$ is the number of branches in a unrooted phylogenetic tree.

In \Table{\ref{tbl: c_vs_pc}}, 
correlation coefficients ($r_{\Delta^s_i \Delta^s_j}$) and
partial correlation coefficients ($r_{\prod_{\perp} \Delta^s_i \prod_{\perp} \Delta^s_j}$) of
substitution probabilities along branches
between sites are classified into four categories; 
significantly positive, positive but insignificant, negative but insignificant, significantly negative.
In addition, sites pairs in each category are classified according to whether
they are contact residue pairs in the protein 3D structure.
Contact residue sites are arbitrarily defined as residue pairs whose minimum atomic distances
are shorter than 5\AA, and which are separated by 6 or more residues along a peptide chain.
The upper table
shows results for Pearson's correlation coefficients and 
the lower table
does those for partial correlation coefficients.
Significantly-positive correlation coefficients are found for almost all site pairs.
In the phylogenetic trees of these protein families
branch lengths are completely heterogeneous.
The expected value of the probability of amino acid substitution in a branch is
an increasing function of branch length;
$\Delta^s_{ib} \approx (1 - \exp( - \mu_i t_b))$ where
$\mu_i$ is an amino acid substitution rate for site $i$.
Thus, Pearson's correlation coefficients of the
expected values of substitution probability over branches between sites
should be significantly positive, as shown in \Table{\ref{tbl: c_vs_pc}}.

On the other hand, a partial correlation coefficient
defined in \Eq{\ref{eq: partial_correlation_coefficient_matrix}}
is a correlation coefficient between residuals 
that cannot be accounted for by a linear regression on 
the vectors of characteristic changes along branches at other sites.
When dependences on other sites in the variation of substitutions
are removed, 
significantly positive correlations
($r > r_t$) are found only in a limited number of site pairs, 
and most site pairs show insignificant correlations.
Furthermore, site pairs in the category of
significantly-positive correlation
tend to be contact residue site pairs with
significantly-high probabilities; see the column of  
positive predictive value, PPV $\equiv \mbox{TP}/(\mbox{TP}+\mbox{FP})$, 
where TP and FP are the numbers of true and false contact residue pairs, respectively.
This result clearly indicates that
the partial correlation coefficients represent
the strength of the direct dependences of substitutions between sites.

\noindent
\subsection*{Co-evolution scores for site pairs}
\vspace*{1em}

Concurrent substitutions between sites require that 
the direct correlation of substitutions
must be positive.  Therefore,
only positive values of the partial correlation coefficients ($\mathcal{C}^s_{ij}$)
are used to define a co-evolution score ($\rho^s_{ij}$) based on concurrent substitutions.  
\begin{eqnarray}
\rho^s_{ij} &\equiv& \max \; (\; \mathcal{C}^s_{ij}, \; 0 \; )
	\label{eq: definition_of_effective_pc_s}
\end{eqnarray}
For all other characteristic variables employed to detect co-evolving site pairs,
the condition of concurrent substitutions between sites are a premise.
Thus, instead of using partial correlations of characteristic variables themselves,
the geometric mean of the partial correlation coefficient of each characteristic variable 
and the co-evolution score based on concurrent substitutions is used as a co-evolution score based on each
characteristic change.
\begin{eqnarray}
\rho^x_{ij} &\equiv&
	\mbox{sgn} \; \mathcal{C}^x_{ij} \;
	( | \rho^s_{ij} \mathcal{C}^x_{ij} | )^{1/2}
	\hspace*{2em} \mbox{for } \hspace*{1em}
	x \in \{ v, c, hb, h, \ldots \}
	\label{eq: definition_of_effective_pc_x}
\end{eqnarray}

As already mentioned in the Method section, 
negative correlations are 
required
for characteristic variables such as
volume, charge, and hydrogen bonding capacity 
to reflect compensatory substitutions.
In \Table{\ref{tbl: pc_each}}, TP and FP for each category of
significantly positive ($\rho^x_{ij} \geq r_t$) 
and negative ($\rho^x_{ij} \leq - r_t$) correlations
under the condition of $|\rho^x_{ij}| \geq \rho^s_{ij} \geq r_t$ 
are listed for each characteristic variable.
In the cases of volume, charge, and hydrogen bonding capacity,
PPV for contact residue pairs is clearly
larger in the category of significantly negative correlation
than significantly positive correlation,
indicating that 
these quantities to detect compensatory substitutions between sites
are good predictors for close spatial proximity.
Besides, there are more site pairs with significantly negative correlations
than with significantly positive correlations, 
clearly indicating the presence of structural constraints against 
these physico-chemical changes.

To improve contact prediction by using characteristic 
variables $\rho^x$ together 
with the characteristic variable $\rho^s$ of concurrent substitutions,
the PPV for the category of significantly positive 
or negative 
correlations
should be larger than the PPV for concurrent substitutions.
Both categories of significantly positive and negative correlations
show better PPVs
in the characteristic variables of hydrophobicity,
$\beta$ and turn propensities, aromatic interaction and branched side-chain.
In the characteristic variables of cross-link capability and ionic side-chain,
only the category of significantly positive correlation
shows better PPV than the category of significantly positive correlation
for concurrent substitutions.
The $\alpha$ propensity is not effective to detect contact residue pairs, 
although it may be effective to detect residue pairs within a helix or
within helices.
Based on these results,
an overall co-evolution score for site pair ($i, j$) is defined here as
\begin{eqnarray}
	\rho_{ij} &\equiv&  \max [
		\rho^s_{ij}, \max( - \rho^v_{ij}, 0), \max( - \rho^c_{ij}, 0),  \max( - \rho^{hb}_{ij}, 0),
		\nonumber
		\\
		& &	\hspace*{2em}
		|\rho^h_{ij}|, |\rho^{\beta}_{ij}|, 
		|\rho^{t}_{ij}|, |\rho^{ar}_{ij}|, 
		|\rho^{br}_{ij}|, \max(\rho^{cl}_{ij}, 0), \max(\rho^{ion}_{ij}, 0) ]
\end{eqnarray}

\noindent
\subsection*{Contact prediction based on the overall co-evolution score $\rho_{ij}$}
\vspace*{1em}

Co-evolving sites pairs are selected for contacts in the decreasing order of
the overall co-evolution score $\rho_{ij}$.
Although this score for co-evolution appears to be able to 
predict contact site pairs, 
preliminary results of contact prediction indicate
that both terminal sites in  multiple sequence alignments often have
large values of $\rho^{x}_{ij}$ ($x \neq s$) for any other site, and
also that there are a few sites
showing extremely large values for $\sum_j H(\rho_{ij} - r_t)$;
the $H$ denotes the Heaviside step function.
Such an anomalous feature probably indicates a poor quality
at these sites in multiple sequence alignments.
Thus, in contact prediction, 
\begin{enumerate}
\item
the co-evolution scores of $\rho^x_{ij}$ ($x \neq s$)
are ignored for both terminal sites in multiple sequence alignments;
that is, $\rho_{ij} \equiv \rho^s_{ij}$.
\item
Also, 
if $\sum_j H(\rho_{ij} - r_t) > 15$,
$\rho_{ij} \equiv \rho^s_{ij}$ will be used for site $i$, and
\item if $\sum_j H(\rho^s_{ij} - r_t) > 15$, $\rho_{ij} \equiv 0$ will be used 
and such a site will be excluded in contact prediction.
\end{enumerate}
The threshold value $r_t$ used here is the value 
of correlation coefficient
corresponding to $\mbox{E-value} = 0.0001$ in the Student's t-distribution.
In the present contact prediction,
only one site that is the N-terminal site in the multiple sequence alignment for the KH\_1
was excluded as an anomalous site.

Needless to say, the norm of any characteristic change vector is almost zero 
for invariant sites; $\| \VEC{\Delta}_{i} \| \simeq 0$. Therefore,
invariant sites are excluded in the present method for contact prediction.

\noindent
\subsection*{Accuracy of contact site pairs predicted on the basis of 
the overall co-evolution score}
\vspace*{1em}

Accuracies of predictions
based on the overall co-evolution score and on the direct information (DI) score\CITE{MCSHPZS:11}
calculated by a maximum entropy model
are compared by using three measures
in \Table{\ref{tbl: prediction_accuracy}} for protein families listed in
\Table{\ref{tbl: protein_families}}.
Those three measures are PPV, mean Euclidean distance from predicted site pairs 
to the nearest true contact (MDPNT) in the 2-dimensional sequence-position space,
and the mean Euclidean distance from every true contact to the nearest predicted site pair 
(MDTNP).
The MDPNT and MDTNP, 
which were defined and used in \CITE{MCSHPZS:11},
are qualitative measures of false positives
and of the spread of predicted site pairs over true contacts, respectively.
Smaller values of these measures indicate better predictions.
For the predictions based on the DI score, 
the filters based on a secondary structure prediction 
and on disulfide bonds are not applied to the DI scores but only the filter\CITE{MCSHPZS:11}
based on the degree of residue conservation at each site is,
in order to compare the prediction accuracies of co-evolution and DI scores themselves.

The reliability of predicted co-evolving site pairs
decreases with decreasing value of co-evolution score, and
co-evolving site pairs are selected in the decreasing order of co-evolution score,
therefore prediction accuracy tends to decrease as the total number of predicted 
sites pairs increases; see \Fig{\ref{fig: PPV_vs_nc_over_tnc}}.
In \Table{\ref{tbl: prediction_accuracy}}, the results of predictions in which the numbers of predicted contacts 
are equal to one fourth or one third of the number of true contacts are listed for each protein family.
One third of the total number of true contacts is equal to the sequence length in the case of Trypsin, which
has the largest number of contacts per residue, and
equal to half of the sequence length in the case of Trans\_reg\_c and 7tm\_1, which
have the smallest number of contacts per residue.
This ratio was chosen, because Marks et al. \CITE{MCSHPZS:11} reported that
one needs about 0.5 to 0.75 predicted distance constraints per residue 
or about 0.25 to 0.35 of the total number of contacts, to achieve
reasonable three-dimensional structure prediction.

In \Fig{\ref{fig: cmap_for_selected_proteins}} and \SFig{\ref{fig: cmap_for_all_proteins}}, 
co-evolving site pairs are shown in the lower half of each triangular map in comparison with 
residue pairs whose minimum atomic distances are less than 5 \AA.
For comparison, contact residue pairs predicted with high DI scores \CITE{MCSHPZS:11}
are also shown in the upper half of each triangular map.
Gray filled-squares, red and indigo filled-circles indicate such residue-residue proximities,
true and false positives in 
contact prediction, respectively. 
It should be noted here that 
residue pairs separated by 5 or fewer positions ($|i - j| \leq 5$) in a sequence
may be shown with the gray filled-squares but 
are excluded in the prediction of co-evolving site pairs and
in the contact prediction with the DI score.
The total number of predicted site pairs is equal to one third of
the total number of true contacts in each protein structure.

In \Table{\ref{tbl: prediction_accuracy}}, 
which method is better in the accuracy of contact prediction is indicated
by a bold font.
The PPVs of the present method are comparable to or better than 
the DI method except CH, Kunitz\_BPTI, Lectin\_C, and RNase\_H.
In \Fig{\ref{fig: PPV_vs_nc_over_tnc}}, 
the PPVs of the present method and the DI method
are drawn by solid and dotted lines as a function of 
the ratio of predicted to true contacts, respectively.
Also, the values of MDPNT and of MDTNP are compared between the present and DI methods
and also between the protein families 
in \Figs{\ref{fig: MDPNT_vs_nc_over_tnc} and \ref{fig: MDTNP_vs_nc_over_tnc}},
respectively.
The good performance of the present method is shown
over a wide range of predicted site pairs.

\noindent
\subsection*{Dependence of the prediction accuracy on the number of predicted site pairs}
\vspace*{1em}

The dependences of the accuracy of predicted contacts on their number
are shown in \Fig{\ref{fig: PPV_vs_nc_over_tnc}} for PPV, 
in \SFig{\ref{fig: MDPNT_vs_nc_over_tnc}} for MDPNT, and
in \SFig{\ref{fig: MDTNP_vs_nc_over_tnc}} for MDTNP.
The total number of predicted site pairs takes every 10
from 10 to a sequence length; also
accuracies for the numbers of predicted contacts equal to one third or one fourth
of true contacts are plotted.
Here, in order to compare prediction accuracies between protein families,
the total number of predicted contacts is shown in the scale of
the ratio of predicted to true contacts. 
It is clearly shown that there is an overall trend for PPV to decrease 
monotonically as increasing number of predicted site pairs.
However, exceptional increases of PPV are also observed 
with increasing number of site pairs predicted.
In the protein family of CH, PPV changes from 0.43 to 0.5 as 
the number of predicted site pairs increases from 30 to 50. 
Because except the case of CH such abnormal increases of PPV often occur
in the range of small numbers of predicted site pairs, i.e., from 10 to 30,
they may be caused by statistical fluctuations.

It is shown in \Table{\ref{tbl: prediction_accuracy}} 
and \Fig{\ref{fig: MDPNT_vs_nc_over_tnc}} that
the relationships of MDPNT with the ratio of predicted to true contacts are
almost inverse of that of PPV, indicating that the MDPNT and 
PPV are two different measures of
the quality of predicted site pairs but result in similar evaluations.
On the other hand,
MDTNP, which measures the spread of predicted site pairs over true contacts, 
differently measures the qualities of predicted contacts from PPV and MDPNT.
It tends to decrease monotonically as the increasing number of predicted site pairs
irrespective of the quality of prediction accuracy,
and therefore it is not appropriate to measure the dependence of prediction accuracy on
the total number of predicted site pairs.

\noindent
\subsection*{Dependence of the prediction accuracy on protein fold types}
\vspace*{1em}

As expected, prediction accuracy is different between proteins.  
However, it is unexpected that prediction accuracy may be 
slightly lower for $\alpha$ proteins, at least for the present three proteins, than for $\beta$ proteins; 
see \Fig{\ref{fig: PPV_vs_nc_over_tnc}}.
Especially the prediction accuracy for the membrane protein 7tm\_1 is 
remarkably
lower than other two $\alpha$ proteins.
This feature is observed in both the present and DI methods.
Thus, this feature may originate in differences  
between structural constraints in $\alpha$-$\alpha$ packing
and in packings of $\beta$ strands and of $\beta$ sheets, 
although the low prediction accuracy for the membrane protein 7tm\_1 
would
result from $\alpha$-$\alpha$ packing peculiar to membrane proteins.
Here it should be noted that the $\alpha$ proteins have less contacts per residue than
the $\beta$ proteins; see \Table{\ref{tbl: prediction_accuracy}}.
A definitive answer must be postponed until more $\alpha$ proteins are analyzed.

\noindent
\subsection*{Dependence of the prediction accuracy on the diversity and the number of sequences used}
\vspace*{1em}

Multiple subsets of a full alignment are generated by using different values of threshold
$T_{bt}$ for branch length to remove OTUs connected to their parent nodes with short branches.
In \Fig{\ref{fig: PPV_vs_n_seqs}}, \SFig{\ref{fig: MDPNT_vs_n_seqs}}, and
\SFig{\ref{fig: MDTNP_vs_n_seqs}}, 
the PPVs, MDPNTs, and MDTNPs calculated from each data are plotted 
against the number of sequences used, respectively.
Because the threshold values used to generate each dataset should also affect 
the accuracy of prediction, they are written near each data point.
A general tendency is of course that the PPV and MDPNT are improved by using more sequences.
However,  the number of sequences and the threshold $T_{bt}$ where accuracy improvement is saturated 
are very different between protein families.
For example, in the case of SH3\_1, no significant improvement in the PPV and MDPNT is observed
in a wide range of $0.2 \geq T_{bt} \geq 0.001$, even if the number of sequences increases from 1500 to 4000. 
In RNase\_H,  the PPV and MDPNT are almost constant in the range of $0.05 \geq T_{bt} \geq 0.001$ 
and $2120 \leq n_{\script{otu}} \leq 7048$.
In Response\_reg, 
after the PPV reaches the highest value 0.73 at $T_{bt} = 0.6$ and 
$n_{\script{otu}} = 3344$, it even decreases to 0.69 in $3344 \leq n_{\script{otu}} \leq 7613$, 
although its decrement is not large and the MDPNT is almost constant in this region.
Multiple sequence alignments may include many sites where significant fraction of sequences have
deletions, reducing effectively the number of sequences; for example, RNase\_H.
However, it may be worth increasing the number of sequences 
until $T_{bt} \approx 0.01$. Here calculations have been carried out
until $n_{\script{otu}} \approx 7000$ or $T_{bt} \approx 0.01$.
A few thousand sequences are necessary to get a reliable prediction 
for proteins consisting of a few hundred residues.

Some data points in \Fig{\ref{fig: PPV_vs_n_seqs}},  
\SFig{\ref{fig: MDPNT_vs_n_seqs}}, and \SFig{\ref{fig: MDTNP_vs_n_seqs}}
correspond to datasets generated by using the same value of threshold
but by removing different OTUs.
PPV often differs between such datasets, although the difference of PPV 
ranges from a few percent to 8 percent; see 
the PPVs for $T_{bt}=0.2$ of Trans\_ref\_C,
$T_{bt}=0.02$ of CH, and $T_{bt}=0.5$ of Cadherin in \Fig{\ref{fig: PPV_vs_n_seqs}}.
This fact indicates that the distribution of sequences in a sequence space
significantly affects prediction accuracy.
Also, it is indicated that
some site pairs predicted are still based on rare events of concurrent substitutions in a tree.




\section*{Discussion}
\subsection*{Significance of compensatory substitutions in protein evolution}

It has been shown that site pairs giving the significant values of
partial correlation coefficients for substitutions,
which concurrently occurred in branches of a
phylogenetic tree and would be mostly compensatory substitutions,
well correspond to contact site pairs in protein 3D structures.
In compensatory substitutions, the fitness of first mutations must be negative,
and successive mutations must occur to compensate the negative effect of the first mutation.
A time scale in which compensatory mutations successively occur is much shorter than 
the time scale of protein evolution that is the order of fixation time for neutral mutations, otherwise
negative mutants will be eliminated from a gene pool by selection.
Thus, negative substitutions and their compensatory substitutions are expected to be observed 
as concurrent substitutions in the same branch of a tree.
If substitutions are completely neutral,  
there will be no correlation in time when substitutions occur.
Thus, a fact that contact site pairs can be well predicted by the present method 
indicates that compensatory substitutions are significant in protein evolution.
Significance of compensatory substitutions was also indicated by 
a fact that likelihoods of phylogenetic trees
can be significantly improved by taking account of codon substitutions 
with multiple nucleotide changes \CITE{M:11a,M:11b}.

\subsection*{Limitations in prediction accuracy}

Prediction accuracy of contact residue pairs 
is different between the protein families.
Possible reasons for false positives include
(1) statistical noise due to an insufficient
number of sequences, insufficient diversity
of sequences, and incorrect matches in a multiple sequence alignment
and an incorrect phylogenetic tree,
(2) structural and functional constraints from 
other residues, which are not taken into account in 
the calculation of partial correlation coefficients from
a correlation matrix, within a protein or even through other molecules involved in
a molecular complex such as oligomerization, protein-substrate, and protein-DNA,
(3) structural variance in homologous proteins, although each Pfam family 
is assumed to be iso-structural.

In reducing statistical noise, the diversity of sequences in protein families 
is more effective than the number of sequences itself. 
Also, the presence of many deletions in sequences reduces the value of including 
those sequences.
The present subset ($T_{bt}=0.01$) of the full alignment of RNase\_H family 
consists of more than 4700 sequences, but their multiple sequence alignment includes
many sites where the significant fraction of sequences have deletions.
Here, branch lengths of sequences from their parent nodes in a phylogenetic tree
are used to get less sequences but as diverse sequences as possible. 
Sequences whose branch lengths from their parent nodes are longer than 0.01 
amino acid substitutions per site in a NJ tree are needed to be 
more than a few thousand, in order to get useful numbers ($> 35 \%$) of PPV.
This requirement seems to be similar to that 
for  the maximum entropy model\CITE{MCSHPZS:11}, in which
the order of one thousand sequences are required to reduce statistical noises
including phylogenetic bias in frequency counts.

\subsection*{Dependence on the accuracies of sequence alignment and phylogenetic tree }

We do not extensively examine the dependence of contact prediction on the accuracy of 
phylogenetic tree.  We tried to optimize a phylogenetic tree as well as model parameters
for Kunitz\_BPTI, Lectin\_C, RNase\_H, and Response\_reg
before evaluating characteristic variables at each site in each branch.
The prediction accuracy of contact site pairs were not significantly improved.
The improvement of prediction accuracy by a tree optimization from a NJ tree
was within a few percent.
The optimization of tree may be not cost-effective, because
it requires a intensive calculation especially for 
thousands of sequences.  
An accurate multiple sequence alignment would be more critical to increase prediction accuracy.

In the present calculations,
sites that are 
deletions
in a target protein structure are excluded in 
the optimization of tree branches and in the calculation of partial correlation coefficients.
The calculation of partial correlation coefficients by including those sites has been attempted
for the Kunitz\_BPTI and RNase\_H domain families.
No improvement was obtained at least for those protein families.

\subsection*{A difference from methods of extracting residue type dependencies between sites}

So far remarkable improvements in the accuracy of contact prediction were all achieved
by extracting essential correlations of amino acid type between residue positions
from multiple sequence alignments\CITE{BN:08,MPBMSZOHW:11,MCSHPZS:11,TS:11}.
Here, almost comparable accuracy of contact prediction has been 
achieved by evaluating direct correlations of concurrent 
and compensatory substitutions between sites. 
The present method cannot be applied to the cases in which
all substitutions are nearly neutral.
In such a case, structural and functional constraints from
closely-located sites in protein 3D structures 
are reflected only in the joint distribution of amino acid types between the sites. 

Residue-residue interactions maintaining secondary structures 
appear to be more easily detected 
by the joint distribution of amino acid types between the sites
than concurrent substitutions.  
In general, the present method detects less dependences between
neighboring sites along a sequence than the other.
Marks et al. \CITE{MCSHPZS:11} reported that residue pairs separated 
by four or five positions in a sequence often have high DI scores without being in close
physical proximity in the folded protein.
Even for site pairs separated by more than five positions,
their method based on the joint distribution of amino acid types 
detected more dependences in $\alpha$ helical regions than the present method; see 
7tm\_1 in \Fig{\ref{fig: cmap_for_selected_proteins}}.

From a such viewpoint, methods of 
extracting direct correlations of amino acid type between sites
may be better 
for extracting direct dependences between sites 
than those of detecting compensatory substitutions
in a tree.   
However,  interactions between closely-located sites do not necessarily  
result in distinct correlations of amino acid type between the sites.
For example, hydrophobic interactions are relatively non-specific, but
significantly contribute to residue-residue interactions inside protein structures.
The types of amino acids found inside protein structures are mainly hydrophobic.
In the case of membrane proteins,  most of amino acids embedded in membrane are hydrophobic.
In the case that residue-residue interactions do not result in distinct correlations of amino acid type 
between sites, methods of detecting compensatory substitutions should perform better than
methods of extracting direct correlations of amino acid type between sites.
Membrane proteins may be this case; see 7tm\_1 in \Fig{\ref{fig: PPV_vs_nc_over_tnc}}.
Structural analyses of membrane proteins, especially 
the determinations of protein coordinates in transmembrane regions,
are difficult in comparison of globular proteins.  The present method for contact prediction
could facilitate the coordinates determinations of membrane proteins.

Besides,
the observed joint distributions of amino acid types include more or less phylogenetic bias as
a statistical noise, but there is no such a noise in the present method.
Thus, the both types of methods are complementary to each other.

\subsection*{A difference between Bayesian graphical models and the present model}

A Bayesian graphical model was applied to
disentangling direct from indirect dependencies between residue positions
in multiple sequence alignments of proteins \CITE{BN:08}.
In the Bayesian graphical model, an acyclic directed graph is
assumed for site dependence, 
although interactions between sites in protein structures act on each other.
A causal relationship between substitutions is of course directional.
However,
substitutions at a site affect on closely-located sites, and
also the site is affected by substitutions at those surrounding sites.
Thus, dependence between sites should be assumed to be bidirectional or undirectional.
Unlike Bayesian graphical models, 
an undirected graph is assumed in a Gaussian graphical model \CITE{E:00}, in which
a null edge between two nodes encodes that 
random variables assigned to the nodes are 
conditionally independent of each other given the values of random variables 
assigned to other nodes.  
Assuming that a joint probability density distribution of random variables is
a multivariate Gaussian distribution, two random variables  are
conditionally independent given the values of other random variables 
if and only if a partial correlation coefficient between the two random variables
is equal to zero.
Thus, the present model based on partial correlation coefficients can be
regarded as a Gaussian graphical model in which an undirected graph is assumed 
for dependences between sites and a feature vector $\VEC{\Delta}_i$ is assigned to node $i$ 
as the observed values of a random variable.
This is one of essential differences between the present model and 
the Bayesian models\CITE{PLFP:08,BN:08,BN:10},
although there is another essential difference that the joint distributions of
residues at sites were analyzed in \CITE{BN:08,BN:10}.


\section*{Methods}

\noindent
\subsection*{Mean of characteristic changes accompanied by substitutions at each site in each branch of a phylogenetic tree
in a maximum likelihood model}
\vspace*{1em}

Assuming that substitutions occur independently at each site,
a likelihood $P(\mathcal{A} | T, \Theta)$ of a sequence alignment $\mathcal{A}$ 
in a phylogenetic tree $T$ under a evolutionary model $\Theta$ is represented as a
product over sites of the likelihood of a sequence alignment $\mathcal{A}_i$ for site $i$.
\begin{eqnarray}
	P(\mathcal{A} | T, \Theta) &=& \prod_i P(\mathcal{A}_i | T, \Theta)
		\\
	P(\mathcal{A}_i | T, \Theta) &=& \sum_{\theta_{\alpha}} P(\mathcal{A}_i | T, \Theta, \theta_{\alpha}) P(\theta_{\alpha})
\end{eqnarray}
where the distribution of $\theta_{\alpha}$ for 
the variation of selective constraint
\CITE{M:11a,M:11b} are assumed to be \textit{a priori} equal to $P(\theta_{\alpha})$,
The evolutionary model $\Theta$ corresponds to a substitution model of amino acid or codon.
Then, if substitutions are assumed to be in the equilibrium state of a time-reversible Markov process,
the likelihood of a sequence alignment $\mathcal{A}_i$ for site $i$ will be calculated
by taking any node as a root node. Let us assume here that the root node is a left node ($v_{bL}$) of a branch $b$.
\begin{eqnarray}
	P(\mathcal{A}_i | T, \Theta, \theta_{\alpha}) &=& 
		\sum_{\kappa} \sum_{\lambda} 
		P(\mathcal{A}_i | v_{bL} = \kappa, v_{bR} = \lambda, T, \Theta, \theta_{\alpha}) 
\end{eqnarray}
\begin{eqnarray}
\lefteqn{
	P(\mathcal{A}_i | v_{bL} = \kappa, v_{bR} = \lambda, T, \Theta, \theta_{\alpha}) \equiv
}
	/nonumber \\
	& &
	P_{bL}(\mathcal{A}_i | v_{bL} = \kappa, T, \Theta, \theta_{\alpha}) 
		f_{\kappa} P(\lambda | \kappa, t_b, \Theta, \theta_{\alpha})
		P_{bR}(\mathcal{A}_i | v_{bR} = \lambda, T, \Theta, \theta_{\alpha}) 
\end{eqnarray}
where $\kappa$ and $\lambda$ indicate the type of codon or amino acid depending on the evolutionary model,
and $f_{\kappa}$ is the equilibrium frequency of $\kappa$.
$P(\lambda | \kappa, t_b, \Theta, \theta_{\alpha})$ is a substitution probability from $\kappa$ to $\lambda$
at the branch $b$ whose length is equal to $t_b$.
$P_{bL}(\mathcal{A}_i | v_{bL} = \kappa, T, \Theta, \theta_{\alpha})$ is
a conditional likelihood of the left subtree with $v_{bL} = \kappa$ \CITE{F:81}.
In the maximum likelihood (ML) method for phylogenetic trees, the tree $T$ and 
parameters $\Theta$ are estimated by maximizing the likelihood.
\begin{eqnarray}
	(\hat{T}, \hat{\Theta}) &=& \arg \max_{T, \Theta} P(\mathcal{A} | T, \Theta)
\end{eqnarray}
In this model, 
the mean $\Delta_{ib}$ of a quantity 
$\Delta_{\kappa\lambda}$ 
accompanied by substitutions from $\kappa$ to  $\lambda$
at each site $i$ in each branch $b$ can be calculated as follows.
\begin{eqnarray}
	\Delta_{ib}(\mathcal{A}_i, \hat{T}, \hat{\Theta}, \theta_{\alpha}) &\equiv& \sum_{\kappa, \lambda} 
		\frac{
		\Delta_{\kappa, \lambda} 
		P(\mathcal{A}_i | v_{bL} = \kappa, v_{bR} = \lambda, \hat{T}, \hat{\Theta}, \theta_{\alpha}) 
		}
		{
		P(\mathcal{A}_i | \hat{T}, \hat{\Theta}, \theta_{\alpha}) 
		}
		\\
	\Delta_{ib}(\mathcal{A}_i, \hat{T}, \hat{\Theta})
		&=& \sum_{\theta_{\alpha}} \Delta_{ib}(\mathcal{A}_i, \hat{T}, \hat{\Theta}, \theta_{\alpha}) 
			P(\theta_{\alpha} | \mathcal{A}_i, \hat{T}, \hat{\Theta})
		\label{eq: posterior_mean_of_delta}
\end{eqnarray}
where $P(\theta_{\alpha} | \mathcal{A}_i, \hat{T}, \hat{\Theta})$ is a posterior probability calculated from
\begin{eqnarray}
	P(\theta_{\alpha} | \mathcal{A}_i, \hat{T}, \hat{\Theta}) &=&
		\frac{
			P(\mathcal{A}_i | \hat{T}, \hat{\Theta}, \theta_{\alpha}) P(\theta_{\alpha}) 
		}
		{
			P(\mathcal{A}_i | \hat{T}, \hat{\Theta})
		}
\end{eqnarray}

If $\Delta_{\kappa\lambda}$ is defined to be equal to $1$ for $\kappa \neq \lambda$
and $0$ for $\kappa = \lambda$, 
$\Delta_{ib}(\mathcal{A}_i, \hat{T}, \hat{\Theta})$
will represent the expected value of substitution probability
at site $i$ in branch $b$.
Let us define a vector $\VEC{\Delta}_i$ as follows, and 
consider the correlation of the two vectors, 
$\VEC{\Delta}_i$ and $\VEC{\Delta}_j$.
\begin{eqnarray}
	\VEC{\Delta}_{i} &\equiv& ( \ldots \; , \; \Delta_{ib}(\mathcal{A}_i, \hat{T}, \hat{\Theta}) - \frac{\sum_b \Delta_{ib}(\mathcal{A}_i, \hat{T}, \hat{\Theta})}{ \sum_b 1} \; , \; \ldots )^{\prime}
\end{eqnarray}
where $\prime$ denotes the transpose of a matrix.
A correlation matrix 
$C$ 
is defined to be a matrix 
whose $(i, j)$ element is the correlation coefficient 
$r_{\Delta_i \Delta_j}$
between $\VEC{\Delta}_i$ and $\VEC{\Delta}_j$.
\begin{eqnarray}
	C_{ij} &\equiv& r_{\Delta_i \Delta_j} = 
		\frac{
			(\VEC{\Delta}_{i}, \VEC{\Delta}_j)
		}
		{
			\|\VEC{\Delta}_i\| \|\VEC{\Delta}_j\|
		}
\end{eqnarray}
where $(\VEC{\Delta}_{i}, \VEC{\Delta}_j)$ denotes the inner product of the two vectors.
The correlation between sites $i$ and $j$ may be an indirect correlation resulting from 
correlations between sites $i$ and $k$ and between sites $k$ and $j$.
To remove such indirect correlations, partial correlation coefficients are
used here.  
The partial correlation coefficient is a correlation coefficient between
residual vectors
($\Pi_{\bot\{\Delta_{k\neq i, j}\}} \VEC{\Delta}_i$ and $\Pi_{\bot\{\Delta_{k\neq i, j}\}} \VEC{\Delta}_j$) 
of given two vectors that are perpendicular to a subspace
consisting of other vectors except those two vectors ($\VEC{\Delta}_i$ and $\VEC{\Delta}_j$)
and therefore cannot be accounted for
by a linear regression on other vectors;
$\Pi_{\bot\{\Delta_{k\neq i, j}\}}$ is a projection operator to a space perpendicular to the subspace.
If the correlation matrix is regular, then the partial correlation coefficients $\mathcal{C}_{ij}$ will be
related to the $(i, j)$ element of its inverse matrix.
\begin{eqnarray}
	\mathcal{C}_{ij}
		&\equiv&
	r_{\Pi_{\bot\{\Delta_{k\neq i, j}\}} \Delta_i \Pi_{\bot\{\Delta_{k\neq i, j}\}} \Delta_j }
		\equiv
	\frac{ ( \Pi_{\bot\{\Delta_{k\neq i, j}\}} \VEC{\Delta}_i \; , \;  \Pi_{\bot\{\Delta_{k\neq i, j}\}} \VEC{\Delta}_j ) } 
	{ \| \Pi_{\bot\{\Delta_{k\neq i, j}\}} \VEC{\Delta}_i \| \; \| \Pi_{\bot\{\Delta_{k\neq i, j}\}} \VEC{\Delta}_j \| }
 		=
		- \; \frac{ (C^{-1})_{ij} }
			{
			 ( (C^{-1})_{jj} (C^{-1})_{ii} )^{1/2}
			}
		\label{eq: partial_correlation_coefficient_matrix}
\end{eqnarray}

\noindent
\subsection*{Characteristic variables indicating co-evolution between sites}
\vspace*{1em}

The following characteristic changes accompanied by substitutions whose correlations indicate 
co-evolution between sites have been used here.
\begin{enumerate}
\item{Occurrence of amino acid substitution.}

The most primary quantity is one ($\Delta^{s}$) that 
is defined as follows and indicates the occurrence of amino acid
substitution at a site.
\begin{eqnarray}
	\Delta^{s}_{\kappa, \lambda} &\equiv& 1 - \delta_{a_\kappa, a_\lambda}
\end{eqnarray}
where $\delta_{a_\kappa, a_\lambda}$ is the Kronecker's $\delta$ that takes 
1 if $a_\kappa=a_\lambda$ and 0 otherwise.
The $a_\kappa$ is the type of amino acid corresponding to $\kappa$.
$\Delta^s_{ib}(\mathcal{A}_i, \hat{T}, \hat{\Theta})$ 
in \Eq{\ref{eq: posterior_mean_of_delta}} indicates the
expected value of the probability of amino acid substitution at site $i$ in branch $b$.
This quantity was also used\CITE{SKS:94,PLFP:08} for the prediction of contact residue pairs 
in protein structures.

\item{Volume change of a side chain accompanied by an amino acid substitution.}

Protein structures must be tightly packed \CITE{R:77},
and therefore mutations between amino acids whose side chain volumes 
significantly differ tend to unstabilize protein structures and
therefore will be eliminated from a gene pool by selection \CITE{GM:78} unless 
the volume change is compensated by successive mutations at sites closely located in
protein structures.
Thus, the volume changes of side chains caused by amino acid substitutions
are used to detect co-evolution between closely located sites in protein structures.
\begin{eqnarray}
	\Delta^{v}_{\kappa, \lambda} &\equiv& \mbox{side\_chain\_volume}_{a_\lambda} - \mbox{side\_chain\_volume}_{a_\kappa}
\end{eqnarray}
where $\mbox{side\_chain\_volume}_{a_\lambda}$ means the volume of 
side chain $a_\lambda$. The amino acid volumes used here are the mean volume 
occupied by each type of amino acid in protein
structures, and taken from the set named BL+ in Table 6 of \CITE{TTCG:99}; 
the volume of a half cystine (labeled as ''cys'' in the table) is used here for a cysteine.

\item{Charge change of a side chain accompanied by an amino acid substitution.}

Charge-charge interactions in protein structures are known to be significant.
Substitutions that keep favorable charge-charge interactions are expected to be
advantageous in selection.  
\begin{eqnarray}
	\Delta^{c}_{\kappa, \lambda} &\equiv& \mbox{side\_chain\_charge}_{a_\lambda} - \mbox{side\_chain\_charge}_{a_\kappa}
\end{eqnarray}
where $\mbox{side\_chain\_charge}_{a_\kappa}$ represents 
a charge of side chain type $a_\kappa$
and takes $1$ for positively charged side chains (arg and lys), 
$0.1$ for his,
and $-1$ for negatively charged ones (asp, glu). 

\item{Change of hydrogen-bonding capability accompanied by an amino acid substitution.}

One of the most important interactions to stabilize protein structures is 
a hydrogen-bonding interaction.  
Substitutions that keep hydrogen-bonds are favorable.
In order to detect whether hydrogen-bonds between side chains can be kept despite
substitutions, the change of hydrogen-bonding capability is defined here as
\begin{eqnarray}
	\Delta^{hb}_{\kappa, \lambda} &\equiv& 
		\mbox{acceptor\_capability}_{a_\lambda} - \mbox{acceptor\_capability}_{a_\kappa}
		+ 
		/nonumber \\
	& & \hspace*{2em}
	\mbox{donor\_capability}_{a_\lambda} - \mbox{donor\_capability}_{a_\kappa}
\end{eqnarray}
where
$\mbox{acceptor\_capability}_{a_\kappa}$ takes $-1$ if a side chain $a_\kappa$
can be an hydrogen-bonding acceptor and 0 otherwise.  
$\mbox{Donor\_capability}_{a_\lambda}$ takes $1$ if a side chain $a_\lambda$
can be a hydrogen-bonding donor and 0 otherwise.  
Hydrogen-bonding acceptors are asn, asp, gln, glu, his, ser, thr, and tyr.
Hydrogen-bonding donors are arg, asn, gln, his, lys, ser, thr, trp, and tyr.
A negative correlation is expected for this quantity
between closely located sites in a protein 3D structure.

\item{Change of hydrophobicity accompanied by an amino acid substitution}

Also,  hydrophobic interactions are crucial for a polypeptide chain to be folded into 
a unique three-dimensional structure.  Hydrophobic interactions may be
correlated between substitutions at nearby sites in a protein 3D structure.
\begin{eqnarray}
	\Delta^{h}_{\kappa, \lambda} &\equiv& e_{a_\lambda r} - e_{a_\kappa r}
\end{eqnarray}
where $e_{a_\kappa r}$ is the mean contact energy of an amino acid 
$a_\kappa$ with 
surrounding residues ($r$) in protein structures; 
see \CITE{MJ:96} for its exact definition.

\item{Changes of $\beta$ and turn propensities accompanied by an amino acid substitution.}

Changes of $\beta$ and turn propensities \CITE{CF:78}
are also examined.
\begin{eqnarray}
	\Delta^{\beta}_{\kappa, \lambda} &\equiv& \beta\mbox{\_sheet\_propensity}_{a_\lambda} - \beta\mbox{\_sheet\_propensity}_{a_\kappa}
	\\
	\Delta^{t}_{\kappa, \lambda} &\equiv& \mbox{turn\_propensity}_{a_\lambda} - \mbox{turn\_propensity}_{a_\kappa}
\end{eqnarray}
where $\beta\mbox{\_sheet\_propensity}_{a_\kappa}$ is the value of $\beta$ sheet propensity \CITE{CF:78} of amino acid $a_\kappa$.
The change of $\alpha$ propensity is also examined but it is not used as a co-evolution score.

\item{Change of the capability of aromatic interaction accompanied by an amino acid substitution.}

\begin{eqnarray}
	\Delta^{ar}_{\kappa, \lambda} &\equiv& \delta_{\script{aromatic\_side\_chains},a_\lambda} - \delta_{\script{aromatic\_side\_chains},a_\kappa}
\end{eqnarray}
where $\delta_{\script{aromatic\_amino\_acids},a_\kappa}$ is equal to 1 if $a_\kappa$ is 
one of aromatic side-chains (his, phe, trp, and tyr) and 0 otherwise.

\item{Change of branched side-chain accompanied by an amino acid substitution.}

\begin{eqnarray}
	\Delta^{br}_{\kappa, \lambda} &\equiv& \delta_{\script{aliphatic\_branched\_side\_chains},a_\lambda} - \delta_{\script{aliphatic\_branched\_side\_chains},a_\kappa}
\end{eqnarray}
where $\delta_{\script{branched\_side\_chains},a_\kappa}$ is equal to 1 if $a_\kappa$ is one of aliphatic branched side-chains (ile, leu and val), 
and 0 otherwise.

\item{Change of cross-link capability accompanied by an amino acid substitution.}

\begin{eqnarray}
	\Delta^{cl}_{\kappa, \lambda} &\equiv& \delta_{\script{cross\_link},a_\lambda} - \delta_{\script{cross\_link},a_\kappa}
\end{eqnarray}
where $\delta_{\script{cross\_link},a_\kappa}$ is equal to 1 if $a_\kappa$ is one of asn, gln, ser and thr, and 0 otherwise.

\item{Change of ionic side-chain accompanied by an amino acid substitution.}

\begin{eqnarray}
	\Delta^{ion}_{\kappa, \lambda} &\equiv& \delta_{\script{inonic\_side\_chains},a_\lambda} - \delta_{\script{inonic\_side\_chains},a_\kappa}
\end{eqnarray}
where $\delta_{\script{ionic\_side\_chains},a_\kappa}$ is equal to 1 if $a_\kappa$ is one of inonic side-chains (asp, glu, arg, and lys), 
0.1 if $a_\kappa$ is his, and 0 otherwise.

\end{enumerate}

\noindent
\subsection*{A mechanistic codon substitution model for 
the maximum likelihood inference of phylogenetic tree}
\vspace*{1em}

A mechanistic codon substitution model, in which
each codon substitution rate is 
proportional to the product of
a codon mutation rate and the average fixation probability 
depending on the type of amino acid replacement, 
has advantages \CITE{M:11a,M:11b} over 
nucleotide, amino acid, and empirical codon substitution models
in evolutionary analysis of protein-coding sequences,
because
mutation at the nucleotide level and selection at the amino acid level
can be separately evaluated.
Even for amino acid sequences of OTUs (operational taxonomic units), 
the mechanistic codon substitution model
with the prior assumption of equal codon usage for 
them yields smaller AIC values (Akaike Information Criterion) than 
any amino acid substitution model does (unpublished).
Thus, the mechanistic codon substitution model \CITE{M:11b}
is used here to evaluate the likelihood of a phylogenetic tree and
the posterior means of characteristic variables at
each site in each branch.

In the mechanistic codon substitution model, in which substitutions are
assumed to be in the stationary state of 
a time-homogeneous reversible Markov process,
the substitution probability matrix in time $t$ 
is represented as $\exp R t$ with a substitution rate matrix $R$,
which is defined as
\begin{eqnarray}
R_{\mu\nu}
        &=& C_{\script{onst}}
        \; M_{\mu\nu} \frac{f_{\nu}}{f_{\nu}^{\script{mut}}} e^{w_{\mu\nu}}
                \; \mbox{ for } \mu \neq \nu
        \label{eq: def_substitution_rate_matrix}
\end{eqnarray}
where $M_{\mu \nu}$ is the mutation rate from codon $\mu$ to $\nu$,
$f_{\nu}^{\script{mut}}$ is the equilibrium frequency of
codon $\nu$ in nucleotide mutations, 
$f_{\nu}$ is the equilibrium codon frequency,
$\frac{f_{\nu}}{f_{\nu}^{\script{mut}}} e^{w_{\mu\nu}}$ is the 
average rate of fixation, and 
$w_{\mu\nu}$ is the selective constraints for mutations from $\mu$ to $\nu$;
refer to \CITE{M:11b} for details.
Assuming that nucleotide mutations occur independently at each codon position but
multiple nucleotide mutations in a codon can instantaneously occur,
the mutation rate matrix $M$ is approximated with
9 parameters;
the ratios of nucleotide mutation rates,
$ m_{tc|ag}/m_{[tc][ag]}$, $m_{ag}/m_{tc|ag}$,
$m_{ta}/m_{[tc][ag]}$, $m_{tg}/m_{[tc][ag]}$, 
and $m_{ca}/m_{[tc][ag]}$,
the relative ratio $m$ of multiple nucleotide changes, 
and the equilibrium nucleotide frequencies
in nucleotide mutations, 
$f_{a}^{\script{mut}}$, $f_{c}^{\script{mut}}$, and $f_{g}^{\script{mut}}$.
The selective constraint $w_{\mu \nu}$ for a protein family 
is approximated 
with a linear function of the mean selective constraints 
that were
evaluated \CITE{M:11a} by ML-fitting a substitution matrix based on 
the mechanistic codon model to 
an empirical amino acid substitution matrix.
Here we use the mean selective constraints
$w_{\mu \nu}^{\script{LG}}$
derived from the empirical amino acid substitution matrix LG \CITE{LG:08}.
The slope $\beta$ and a constant term $w_0$ are parameters;
$w_{\mu \nu} = \beta w_{\mu \nu}^{\script{LG}} + w_0$.
The selective constraint $w_{\mu \nu}$ is
assumed to vary across sites and the variation of selective constraints
\CITE{M:11b}
has been approximated by a discrete gamma distribution\CITE{Y:94} with 4 categories.
Thus, one more parameter is a shape parameter $\alpha$ for the
discrete gamma distribution.
In the result, 12 parameters in addition to the equilibrium frequencies 
of codons must be determined in this model.
See \CITE{M:11b} for full details of these parameters.

The equilibrium frequencies of codons are estimated to be equal to
codon frequencies in sequences of OTUs with the assumption of
equal codon usage for amino acid sequences.
Other 12 parameters were estimated by maximizing the likelihood of
a NJ tree of Pfam seed sequences.
Then, the ML estimates of the parameters obtained from 
the  Pfam seed sequences are used to evaluate
branch lengths and posterior means of characteristic variables
at each site in each branch of NJ trees for the subsets of Pfam full alignments.
NJ trees taken from the Pfam 
were used for the tree topologies, because optimizing tree topologies
for more than a few thousands of sequences
require too much computational time.
Branch optimization of phylogenetic trees and posterior means of 
characteristic variables are calculated using
Phyml\CITE{GG:03} modified for the mechanistic codon substitution model.

\noindent
\subsection*{Definition of contact residue pairs in protein structures}
\vspace*{1em}

Contact residue pairs are arbitrarily defined here as residue pairs
whose minimum atomic distances are shorter than 5 \AA \; and
which are separated by 6 or more residues along a peptide chain.
This definition, especially the latter condition, 
which was used in Marks et al.\CITE{MCSHPZS:11}
is employed here only for the comparison of
the present predictions with their predictions of contact residue pairs. 

The PDB ID of a protein structure used for a target protein in each Pfam family
is listed in \Table{\ref{tbl: protein_families}}.
The amino acid sequences of these PDB entries are just the same as
those of the Uniprot IDs,
which are also listed in \Table{\ref{tbl: protein_families}}.

\newpage

\section*{Acknowledgments}

\bibliography{plos_article}

\clearpage

\renewcommand{\FigureInLegends}[1]{#1}

\section*{Figure Legends}
\renewcommand{\TextFig}[1]{#1}
\renewcommand{\SupFig}[1]{}

\TextFig{

\begin{figure*}[ht]
\FigureInLegends{
\centerline{
\includegraphics*[width=90mm,angle=0]{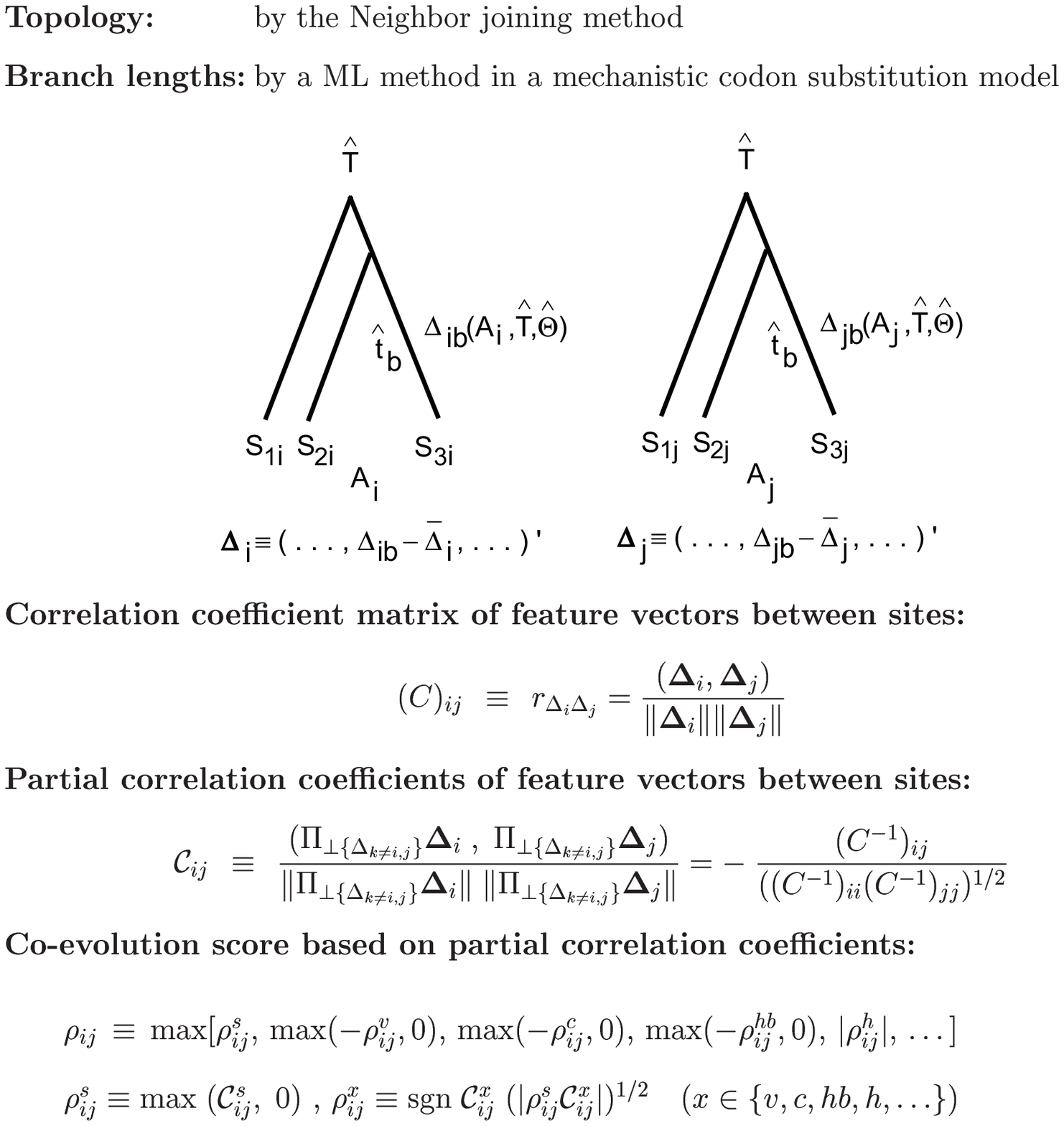}
}
} 
\vspace*{1em}
\caption{
\label{fig: framework}
\BF{Framework of the present model.}
See text for details. \hspace*{20em}
}
\end{figure*}

\FigureInLegends{\newpage}

\begin{figure*}[ht]
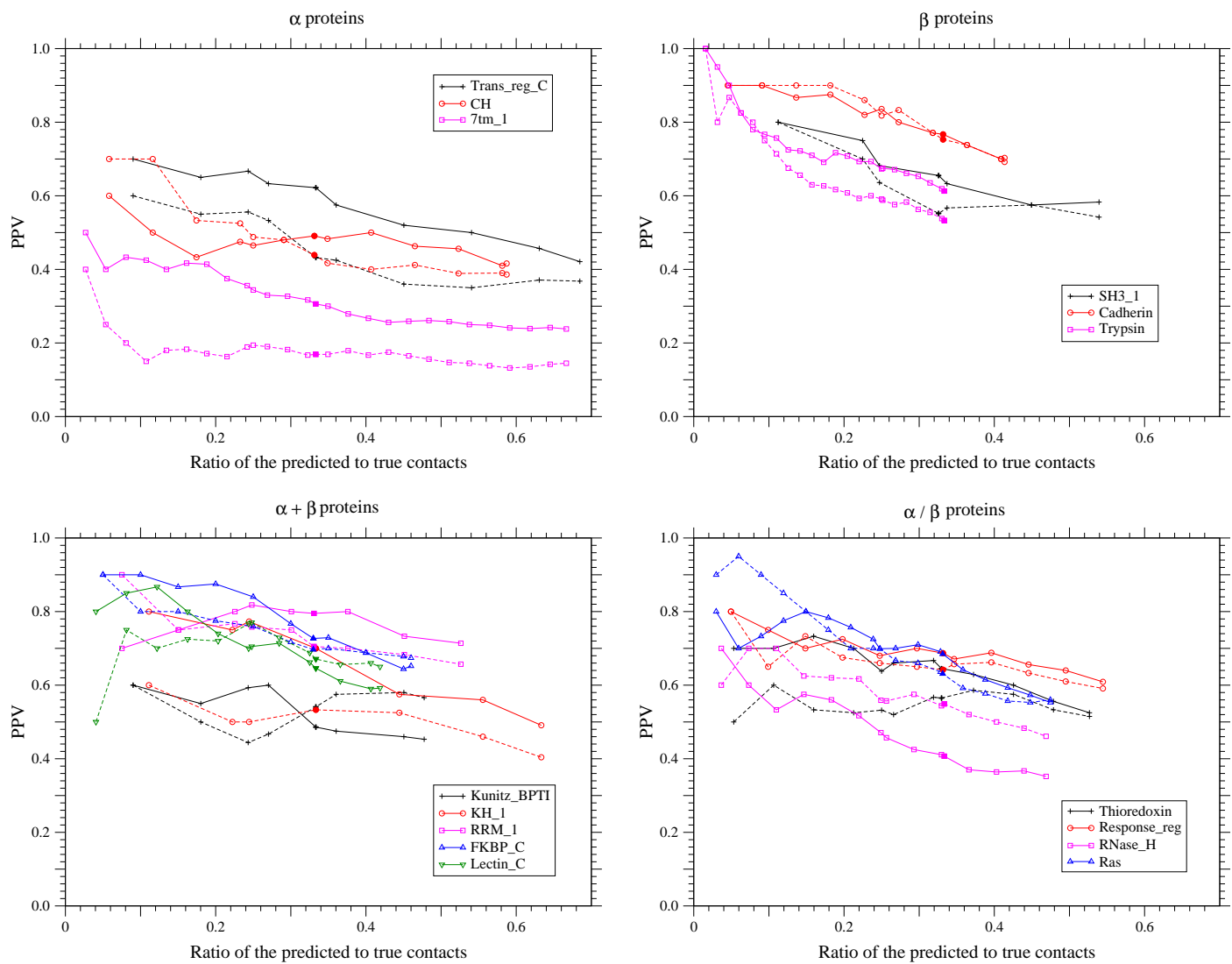

\FigureInLegends{

\centerline{
\includegraphics*[width=90mm,angle=0]{FIGS/PPV_vs_nc_over_tnc.a.eps}
\hspace*{3mm}
\includegraphics*[width=90mm,angle=0]{FIGS/PPV_vs_nc_over_tnc.b.eps}
}
\vspace*{1em}

\centerline{
\includegraphics*[width=90mm,angle=0]{FIGS/PPV_vs_nc_over_tnc.a+b.eps}
\hspace*{3mm}
\includegraphics*[width=90mm,angle=0]{FIGS/PPV_vs_nc_over_tnc.a-b.eps}
}
} 
\vspace*{1em}
\caption{
\label{fig: PPV_vs_nc_over_tnc}
\BF{Dependence of PPV on the number of predicted contacts.}
The dependences of the positive predictive values 
on the total number of predicted contacts
are shown for each protein fold of $\alpha$, $\beta$, $\alpha + \beta$, and $\alpha/\beta$.
The solid and dotted lines show the PPVs of the present method and
the method based on the DI score\CITE{MCSHPZS:11}, respectively.
The total number of predicted site pairs is shown in the scale of
the ratio of the number of predicted site pairs to the number of true contacts.
The total number of predicted site pairs takes every 10
from 10 to a sequence length; also
PPVs for the numbers of predicted site pairs equal to one fourth or one third
of true contacts are plotted.
The filled marks indicate the points corresponding to the number of predicted site pairs
equal to one third of the number of true contacts.
The number of sequences used here for each protein family is one listed in 
\Table{\ref{tbl: protein_families}}.
}
\end{figure*}

\FigureInLegends{\newpage}

\begin{figure*}[ht]
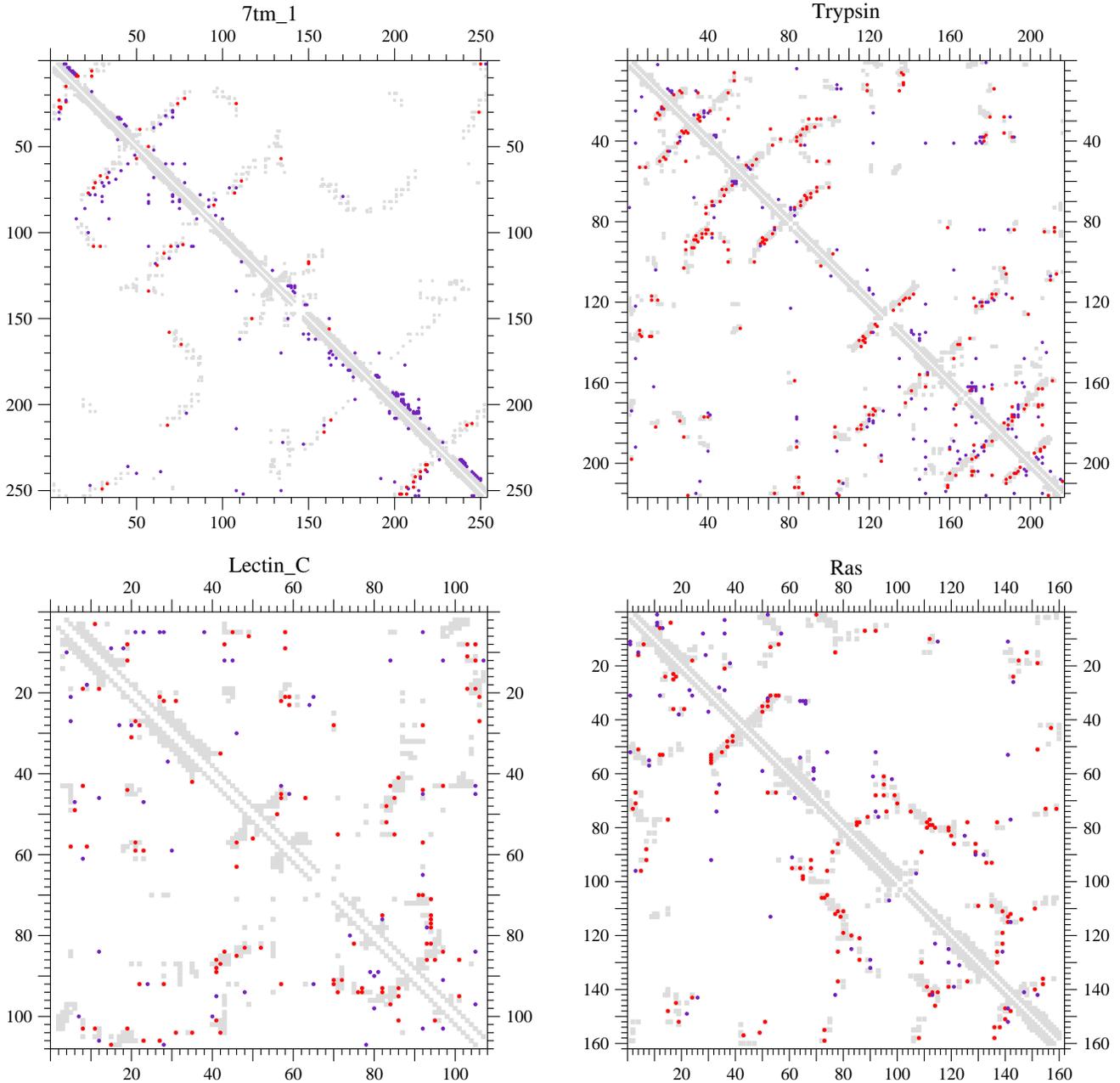

\FigureInLegends{

\centerline{
\includegraphics*[width=83mm,angle=0]{FIGS/7tm_1.cmap.eps}
\hspace*{5mm}
\includegraphics*[width=83mm,angle=0]{FIGS/Trypsin.cmap.eps}
}
\vspace*{1em}

\centerline{
\includegraphics*[width=83mm,angle=0]{FIGS/Lectin_C.cmap.eps}
\hspace*{5mm}
\includegraphics*[width=83mm,angle=0]{FIGS/Ras.cmap.eps}
}
} 
\vspace*{1em}
\caption{
\label{fig: cmap_for_selected_proteins}
\BF{Co-evolving site pairs versus DI residue pairs.}
Residue pairs whose minimum atomic distances are shorter than 5 \AA $\;$
in a protein structure and co-evolving site pairs predicted are shown
by gray filled-squares and by red or indigo filled-circles in the
lower-left half of each figure, respectively.
For comparison, such residue-residue proximities 
and predicted contact residue pairs with high DI scores in \CITE{MCSHPZS:11}
are also shown by gray filled-squares and by red or indigo filled-circles 
in the upper-right half of each figure, respectively; 
only the conservation filter is applied but the filters based on
a secondary structure prediction and for cysteine pairs are not applied to the DI scores.
Red and indigo filled-circles correspond to true and false contact residue pairs, respectively.
Residue pairs separated by five or fewer positions in a sequence may be shown with the
gray filled-squares but are excluded in both the predictions.
The total numbers of co-evolving site pairs and DI residue pairs plotted for each protein 
are both equal to one third of true contacts ($\mbox{TP} + \mbox{FP} = \mbox{\#contacts} / 3$).
The PPVs of both the methods for each protein are listed in \Table{\ref{tbl: prediction_accuracy}}.
}
\end{figure*}

\FigureInLegends{\newpage}

\begin{figure*}[ht]
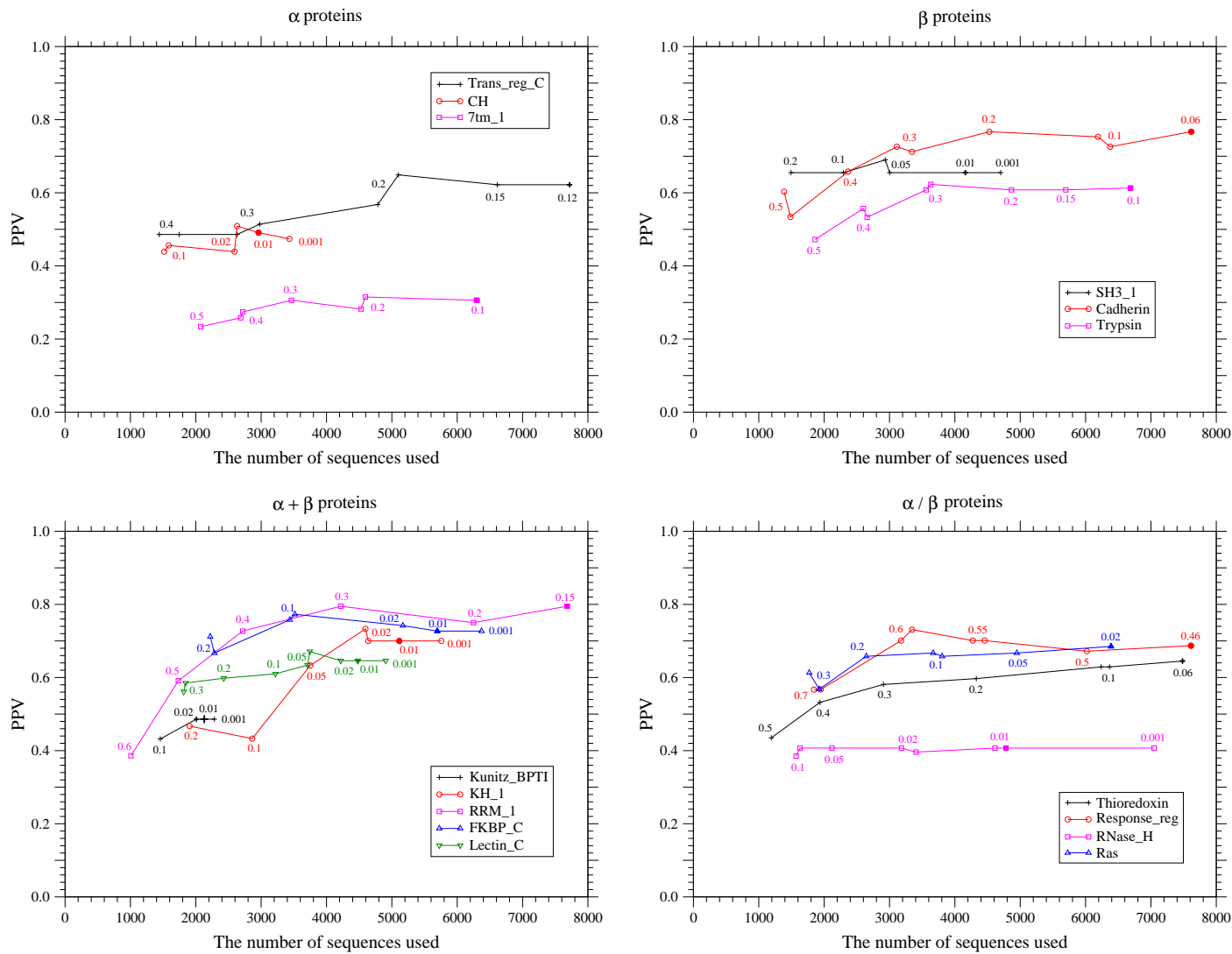

\FigureInLegends{

\centerline{
\includegraphics*[width=90mm,angle=0]{FIGS/PPV_vs_n_seqs.a.eps}
\hspace*{3mm}
\includegraphics*[width=90mm,angle=0]{FIGS/PPV_vs_n_seqs.b.eps}
}
\vspace*{1em}

\centerline{
\includegraphics*[width=90mm,angle=0]{FIGS/PPV_vs_n_seqs.a+b.eps}
\hspace*{3mm}
\includegraphics*[width=90mm,angle=0]{FIGS/PPV_vs_n_seqs.a-b.eps}
}
} 
\vspace*{1em}
\caption{
\label{fig: PPV_vs_n_seqs}
\BF{Dependence of PPV on the number of sequences used.}
The positive predictive values are plotted against
the total number of homologous sequences used
for each prediction.
The filled marks indicate the points corresponding to the number of used sequences
listed for each protein family in \Table{\ref{tbl: protein_families}}.
The values written near each data point indicate the threshold value $T_{bt}$;
OTUs connected to their parent nodes with branches shorter than this
threshold value are removed in the NJ tree of the Pfam full sequences used for each prediction.
Some data points correspond to datasets generated by using the same value of the threshold
but by removing different OTUs.
}
\end{figure*}

} 

\SupFig{

\begin{figure*}[ht]
\vspace*{1em}
\caption{
\label{fig: cmap_for_all_proteins}
\BF{Co-evolving site pairs versus DI residue pairs.}
Residue pairs whose minimum atomic distances are shorter than 5 \AA $\;$
in a protein structure and co-evolving site pairs predicted are shown
by gray filled-squares and by red or indigo filled-circles in the
lower-left half of each figure, respectively.
For comparison, such residue-residue proximities 
and predicted contact residue pairs with high DI scores in \CITE{MCSHPZS:11}
are shown by gray filled-squares and by red or indigo filled-circles 
in the upper-right half of each figure, respectively; 
only the conservation filter is applied but the filters based on
a secondary structure prediction and for cysteine pairs are not applied to the DI scores.
Red and indigo filled-circles correspond to true and false contact residue pairs, respectively.
Residue pairs separated by five or fewer positions in a sequence may be shown with the
gray filled-squares but are excluded in both the predictions.    
The total numbers of co-evolving site pairs and DI residue pairs plotted for each protein
are both equal to one third of true contacts ($\mbox{TP} + \mbox{FP} = \mbox{\#contacts} / 3$).
The PPVs of both the methods for each protein are listed in \Table{\ref{tbl: prediction_accuracy}}.
}
\end{figure*}

\FigureInLegends{\newpage}

\FigureInLegends{

\centerline{
\includegraphics*[width=80mm,angle=0]{FIGS/Trans_reg_C.cmap.eps}
\hspace*{5mm}
\includegraphics*[width=83mm,angle=0]{FIGS/CH.cmap.eps}
}
\vspace*{1em}

\centerline{
\includegraphics*[width=83mm,angle=0]{FIGS/7tm_1.cmap.eps}
\hspace*{5mm}
\hspace*{80mm}
}
} 
\vspace*{1em}

\FigureInLegends{\newpage}

\FigureInLegends{

\centerline{
\includegraphics*[width=80mm,angle=0]{FIGS/SH3_1.cmap.eps}
\hspace*{5mm}
\includegraphics*[width=83mm,angle=0]{FIGS/Cadherin.cmap.eps}
}
\vspace*{1em}

\centerline{
\includegraphics*[width=83mm,angle=0]{FIGS/Trypsin.cmap.eps}
\hspace*{5mm}
\hspace*{80mm}
}
} 
\vspace*{1em}

\FigureInLegends{\newpage}

\FigureInLegends{
\centerline{
\includegraphics*[width=80mm,angle=0]{FIGS/Kunitz_BPTI.cmap.eps}
\hspace*{5mm}
\includegraphics*[width=80mm,angle=0]{FIGS/KH_1.cmap.eps}
}
\vspace*{1em}

\centerline{
\includegraphics*[width=80mm,angle=0]{FIGS/RRM_1.cmap.eps}
\hspace*{5mm}
\includegraphics*[width=80mm,angle=0]{FIGS/FKBP_C.cmap.eps}
}
\vspace*{1em}

\centerline{
\includegraphics*[width=83mm,angle=0]{FIGS/Lectin_C.cmap.eps}
\hspace*{5mm}
\hspace*{80mm}
}
} 
\vspace*{1em}

\FigureInLegends{\newpage}

\FigureInLegends{

\centerline{
\includegraphics*[width=83mm,angle=0]{FIGS/Thioredoxin.cmap.eps}
\hspace*{5mm}
\includegraphics*[width=83mm,angle=0]{FIGS/Response_reg.cmap.eps}
}
\vspace*{1em}

\centerline{
\includegraphics*[width=83mm,angle=0]{FIGS/RNase_H.cmap.eps}
\hspace*{5mm}
\includegraphics*[width=83mm,angle=0]{FIGS/Ras.cmap.eps}
}
} 
\vspace*{1em}

\FigureInLegends{\newpage}

\begin{figure*}[ht]
\FigureInLegends{

\centerline{
\includegraphics*[width=90mm,angle=0]{FIGS/MDPNT_vs_nc_over_tnc.a.eps}
\hspace*{3mm}
\includegraphics*[width=90mm,angle=0]{FIGS/MDPNT_vs_nc_over_tnc.b.eps}
}
\vspace*{1em}

\centerline{
\includegraphics*[width=90mm,angle=0]{FIGS/MDPNT_vs_nc_over_tnc.a+b.eps}
\hspace*{3mm}
\includegraphics*[width=90mm,angle=0]{FIGS/MDPNT_vs_nc_over_tnc.a-b.eps}
}
} 
\vspace*{1em}
\caption{
\label{fig: MDPNT_vs_nc_over_tnc}
\BF{Dependence of MDPNT on the number of predicted contacts.}
The dependences of 
the mean Euclidean distance from predicted site pairs to the nearest true contact in
the 2-dimensional sequence-position space
on the total number of predicted contacts
are shown for each protein fold of $\alpha$, $\beta$, $\alpha + \beta$, and $\alpha / \beta$.
The solid and dotted lines show the MDPNTs of the present method and
the method based on the DI score\CITE{MCSHPZS:11}, respectively.
The total number of predicted contacts is shown in the scale of
the ratio of the number of predicted contacts to the number of true contacts.
The total number of predicted site pairs takes every 10
from 10 to a sequence length; also
MDPNTs for the numbers of predicted contacts equal to one fourth or one third
of true contacts are plotted.
The filled marks indicate the points corresponding to the number of predicted site pairs
equal to one third of the number of true contacts.
The number of sequences used here for each protein family is one listed in 
\Table{\ref{tbl: protein_families}}.
}
\end{figure*}

\FigureInLegends{\newpage}

\begin{figure*}[ht]
\FigureInLegends{

\centerline{
\includegraphics*[width=90mm,angle=0]{FIGS/MDTNP_vs_nc_over_tnc.a.eps}
\hspace*{3mm}
\includegraphics*[width=90mm,angle=0]{FIGS/MDTNP_vs_nc_over_tnc.b.eps}
}
\vspace*{1em}

\centerline{
\includegraphics*[width=90mm,angle=0]{FIGS/MDTNP_vs_nc_over_tnc.a+b.eps}
\hspace*{3mm}
\includegraphics*[width=90mm,angle=0]{FIGS/MDTNP_vs_nc_over_tnc.a-b.eps}
}
} 
\vspace*{1em}
\caption{
\label{fig: MDTNP_vs_nc_over_tnc}
\BF{Dependence of MDTNP on the number of predicted contacts.}
The dependences of
the mean Euclidean distance from every true contact to the nearest predicted site pair
in the 2-dimensional sequence-position space
on the total number of predicted contacts
are shown for each protein fold of $\alpha$, $\beta$, $\alpha + \beta$, and $\alpha / \beta$.
The solid and dotted lines show the MDTNPs of the present method and
the method based on the DI score\CITE{MCSHPZS:11}, respectively.
The total number of predicted site pairs is shown in the scale of
the ratio of the number of predicted site pairs to the number of true contacts.
The total number of predicted site pairs takes every 10
from 10 to a sequence length; also
MDTNPs for the numbers of predicted site pairs equal to one fourth or one third
of true contacts are plotted.
The filled marks indicate the points corresponding to the number of predicted contacts 
equal to one third of the number of true contacts.
The number of sequences used here for each protein family is one listed in 
\Table{\ref{tbl: protein_families}}.
}
\end{figure*}

\FigureInLegends{\newpage}

\begin{figure*}[ht]
\FigureInLegends{

\centerline{
\includegraphics*[width=90mm,angle=0]{FIGS/MDPNT_vs_n_seqs.a.eps}
\hspace*{3mm}
\includegraphics*[width=90mm,angle=0]{FIGS/MDPNT_vs_n_seqs.b.eps}
}
\vspace*{1em}

\centerline{
\includegraphics*[width=90mm,angle=0]{FIGS/MDPNT_vs_n_seqs.a+b.eps}
\hspace*{3mm}
\includegraphics*[width=90mm,angle=0]{FIGS/MDPNT_vs_n_seqs.a-b.eps}
}
} 
\vspace*{1em}
\caption{
\label{fig: MDPNT_vs_n_seqs}
\BF{Dependence of MDPNT on the number of sequences used.}
The mean Euclidean distance from every predicted site pair to the nearest true contact in
the 2-dimensional sequence-position space
is plotted against
the total number of homologous sequences used
for each prediction.
The filled marks indicate the points corresponding to the number of used sequences
listed for each protein family in \Table{\ref{tbl: protein_families}}.
The values written near each data point indicate the threshold value $T_{bt}$;
OTUs connected to their parent nodes with branches shorter than this
threshold value are removed in the NJ tree of the Pfam full sequences used for each prediction.
Some data points correspond to datasets generated by using the same value of the threshold
but by removing different OTUs.
}
\end{figure*}

\FigureInLegends{\newpage}

\begin{figure*}[ht]
\FigureInLegends{

\centerline{
\includegraphics*[width=90mm,angle=0]{FIGS/MDTNP_vs_n_seqs.a.eps}
\hspace*{3mm}
\includegraphics*[width=90mm,angle=0]{FIGS/MDTNP_vs_n_seqs.b.eps}
}
\vspace*{1em}

\centerline{
\includegraphics*[width=90mm,angle=0]{FIGS/MDTNP_vs_n_seqs.a+b.eps}
\hspace*{3mm}
\includegraphics*[width=90mm,angle=0]{FIGS/MDTNP_vs_n_seqs.a-b.eps}
}
} 
\vspace*{1em}
\caption{
\label{fig: MDTNP_vs_n_seqs}
\BF{Dependence of MDTNP on the number of sequences used.}
The mean Euclidean distance from every true contact to the nearest predicted site pair
in the 2-dimensional sequence-position space
is plotted against
the total number of homologous sequences used
for each prediction.
The filled marks indicate the points corresponding to the number of used sequences
listed for each protein family in \Table{\ref{tbl: protein_families}}.
The values written near each data point indicate the threshold value $T_{bt}$;
OTUs connected to their parent nodes with branches shorter than this
threshold value are removed in the NJ tree of the Pfam full sequences used for each prediction.
Some data points correspond to datasets generated by using the same value of the threshold
but by removing different OTUs.
}
\end{figure*}

} 
\clearpage

\section*{Tables}
\renewcommand{\TableInLegends}[1]{#1}
\renewcommand{\TextTable}[1]{#1}
\renewcommand{\SupTable}[1]{}

\TextTable{

\begin{table}[hb]
\caption{\label{tbl: protein_families}
\BF{
Protein families used.
} 
}

\vspace*{2em}
\TableInLegends{

\begin{tabular}{lrrlllc}
\hline
Pfam ID$^a$     & Seed$^b$ & Full$^c$ &	\multicolumn{2}{c}{Target protein domain} &	Fold type 	& No. sites/Length$^f$	\\
		&	&	& Uniprot ID$^d$	& PDB ID$^e$ &		&	\\
\hline
Trans\_reg\_C	& 362	& 35180	& OMPR\_ECOLI/156-232	& 1ODD-A:156-232	& $\alpha$	& 76/77	\\
CH	& 202	& 5756	& SPTB2\_HUMAN/176-278	& 1BKR-A:5-107	& $\alpha$	& 101/103	\\
7tm\_1	& 64	& 26656	& OPSD\_BOVIN/54-306	& 1GZM-A:54-306	& $\alpha$ (tm)$^g$	& 248/253	\\
SH3\_1	& 61	& 8993	& YES\_HUMAN/97-144	& 2HDA-A:97-144	& $\beta$	& 48/48	\\
Cadherin	& 57	& 18808	& CADH1\_HUMAN/267-366	& 2O72-A:113-212 & $\beta$	& 91/100	\\
Trypsin	& 71	& 14720	& TRY2\_RAT/24-239	& 3TGI-E:16-238	& $\beta$	& 212/216	\\
Kunitz\_BPTI	& 151	& 3090	& BPT1\_BOVIN/39-91	& 5PTI-A:4-56	& $\alpha+\beta$	& 53/53	\\
KH\_1	& 399	& 11484	& PCBP1\_HUMAN/281-343	& 1WVN-A:7-69	& $\alpha+\beta$	& 57/63	\\
RRM\_1	& 79	& 31837	& ELAV4\_HUMAN/48-118	& 1G2E-A:41-111	& $\alpha+\beta$	& 70/71	\\
FKBP\_C	& 174	& 11034	& O45418\_CAEEL/26-118	& 1R9H-A:26-118	& $\alpha+\beta$	& 92/93	\\
Lectin\_C	& 44	& 6530	& CD209\_HUMAN/273-379	& 1SL5-A:273-379	& $\alpha+\beta$	& 103/107	\\
Thioredoxin	& 50	& 16281	& THIO\_ALIAC/1-103	& 1RQM-A:1-103	& $\alpha/\beta$ 	& 99/103	\\
Response\_reg	& 57	& 103232	& CHEY\_ECOLI/8-121	& 1E6K-A:8-121	& $\alpha/\beta$	& 110/114	\\
RNase\_H	& 65	& 13801	& RNH\_ECOLI/2-142	& 1F21-A:3-142 & $\alpha/\beta$	& 128/140	\\
Ras	& 61	& 13525	& RASH\_HUMAN/5-165	&	5P21-A:5-165	& $\alpha/\beta$	& 159/161	\\
\hline
\end{tabular}

{
\small{
\vspace*{1em}
\renewcommand{\baselinestretch}{1.0}
$^a$  Pfam release 26.0 (November 2011) was used.
\newline
$^b$  The number of sequences included in the seed alignment of the Pfam.
\newline
$^c$  The number of sequences included in the full alignment of the Pfam.
\newline
$^d$  Target protein domain in the Pfam family.
\newline
$^e$  A protein structure corresponding to the target protein domain.
\newline
$^f$  Unreliable site positions that are represented by the lower case of characters 
in alignments were excluded in the evaluation of prediction accuracy.
\newline
$^g$ Transmembrane $\alpha$.
}
}

} 
\end{table}

\begin{table}[hb]
\caption{\label{tbl: c_vs_pc}
\BF{
Correlation ($r_{\Delta^s_i \Delta^s_j}$) versus 
partial correlation ($r_{\prod_{\perp}\Delta^s_i \prod_{\perp}\Delta^s_j}$) 
coefficients of concurrent substitutions between sites.
} 
}

\vspace*{2em}
\TableInLegends{

\begin{tabular}{lrrrrrrrrrr}
\hline
Pfam ID & $T_{bt}$$^a$	& \#seqs$^a$ & \multicolumn{2}{c}{$C^s_{ij} \geq r_t$$^b$} & \multicolumn{2}{c}{$r_t > C^s_{ij} > 0$}  
				& \multicolumn{2}{c}{$0 > C^s_{ij} > - r_t$} & \multicolumn{2}{c}{$-r_t \geq C^s_{ij}$}	\\
	 & 		& 	 & TP:FP$^c$ & PPV$^d$ & TP:FP & PPV & TP:FP & PPV & TP:FP & PPV	\\
\hline
Trans\_reg\_C	& 0.12	& 7720	& 102:2282 &0.04& 1:30	& 0.03	& 0:0	& --	& 0:0	& --	\\
CH		& 0.01	& 2960	& 167:4226 &0.04& 2:73	&0.03	& 0:2	& 0.0	& 0:0	& --	\\
7tm\_1		& 0.1	& 6302	& 358:28576&0.01& 0:0	& --	& 0:0	& --	& 0:0	& --	\\
SH3\_1		& 0.01	& 4160	& 74:674 &0.10	& 7:60	& 0.10	& 0:5	& 0.0	& 0:0	& --	\\
Cadherin	& 0.06	& 7617	& 214:3333&0.06& 1:46	&0.02	& 0:7	& 0.0	& 0:0	& --	\\
Trypsin		& 0.1	& 6688	& 617:20312&0.03& 0:0	& --	& 0:0	& --	& 0:0	& --	\\
Kunitz\_BPTI	& 0.01	& 2130	& 86:799 &0.10	& 11:48	&0.19	& 0:2	& 0.0	& 0:0	& --	\\
KH\_1		& 0.01	& 5114	& 78:1116&0.07	& 1:41	&0.02	& 0:4	& 0.0	& 0:0	& --	\\
RRM\_1		& 0.15	& 7684	& 119:1839&0.06	& 0:0	& --	& 0:0	& --	& 0:0	& --	\\
FKBP\_C		& 0.01	& 5695	& 199:3445&0.05	& 0:10	& 0.0	& 0:1	& 0.0	& 0:0	& --	\\
Lectin\_C	& 0.01	& 4479	& 234:4319&0.05	& 1:19	&0.05	& 0:0	& --	& 0:0	& --	\\
Thioredoxin	& 0.06	& 7483	& 188:4180&0.04	& 0:3	&0.0	& 0:0	& --	& 0:0	& --	\\
Response\_reg	& 0.46	& 7613	& 202:5266&0.04	& 0:1	& 0.0	& 0:0	& --	& 0:0	& --	\\
RNase\_H	& 0.01	& 4782	& 271:7152&0.04	& 0:5	& 0.0 	& 0:0	& --	& 0:0	& --	\\
Ras		& 0.02	& 6390	& 329:11304&0.03& 0:0	& --	& 0:0	& --	& 0:0	& --	\\
\hline
\hline
Pfam ID & \multicolumn{2}{l}{\#contacts} & \multicolumn{2}{c}{$\mathcal{C}^s_{ij} \geq r_t$$^b$} & \multicolumn{2}{c}{$r_t > \mathcal{C}^s_{ij} > 0$}  
				& \multicolumn{2}{c}{$0 > \mathcal{C}^s_{ij} > - r_t$} & \multicolumn{2}{c}{$-r_t \geq  \mathcal{C}^s_{ij}$}	\\
 & \multicolumn{2}{l}{\hspace*{1ex}}/\#sites$^c$  & TP:FP$^c$ & PPV$^d$ & TP:FP & PPV & TP:FP & PPV & TP:FP & PPV	\\
\hline
Trans\_reg\_C	& 103/75& 1.4 	& 32:57	&0.36	& 59:1584&0.04	& 12:669&0.02	& 0:2	& 0.0	\\
CH		& 169/100&  1.7	& 16:17	&0.48	& 125:2454&0.05	& 28:1828&0.02	& 0:2	& 0.0	\\
7tm\_1		& 366/247& 1.5	& 36:84	&0.30	&263:15695&0.02	&59:12787&0.005	& 0:10	& 0.0	\\
SH3\_1		& 81/46	& 1.8	& 24:17	&0.59	& 46:516&0.08	&11:206	&0.05	& 0:0	& --	\\
Cadherin	& 215/90& 2.4	& 40:8	& 0.83	&132:1519&0.08	&42:1857& 0.02	& 1:2	& 0.33	\\
Trypsin		& 617/210& 2.9	& 115:75&0.61 	& 383:11331 &0.03& 119:8899 &0.01& 0:7	& 0.0	\\
Kunitz\_BPTI	& 105/51& 2.1	& 16:12	&0.57	& 55:575&0.09	&26:262	& 0.09	& 0:0	& --	\\
KH\_1		& 79/55	& 1.4	& 19:15	&0.56	& 50:707&0.07	& 10:438& 0.02	& 0:1	& 0.0	\\
RRM\_1		& 119/68& 1.8	& 45:36	&0.56	& 63:1257&0.05	& 11:546&0.02	& 0:0	& --	\\
FKBP\_C		& 199/91& 2.2	& 66:51	&0.56	& 103:2114&0.05	& 30:1288&0.02	& 0:3	& 0.0	\\
Lectin\_C	& 243/102& 2.4	& 36:13	& 0.73	& 160:2401&0.06	& 39:1923& 0.02	& 0:1	& 0.0	\\
Thioredoxin	& 188/99& 1.9	& 53:61	&0.46	& 109:2677&0.04	& 26:1442&0.02	& 0:3	& 0.0	\\
Response\_reg	& 202/110& 1.8	& 72:87 &0.45	& 101:3182&0.03	& 28:1988&0.01	& 1:10	&0.09	\\
RNase\_H	& 271/127& 2.1	& 37:56	&0.40	& 161:3700&0.04	& 72:3387&0.02	& 1:14	&0.07	\\
Ras		& 329/158& 2.1	& 81:55	&0.60	& 203:6472&0.03	& 44:4768&0.01	& 1:9	&0.10	\\
\hline
\end{tabular}

{
\small{
\vspace*{1em}
$^a$	OTUs connected to their parent nodes with branches shorter than 
the threshold value $T_{bt}$ are removed from each Pfam full alignment, 
and the number of remaining OTUs is listed.
\newline
$^b$	The $r_t$ is a threshold for a correlation coefficient corresponding to
the E-value $E_t = 0.001$ (the P-value $P_t = E_t / n_{\script{pairs}}$)
in the Student's t-distribution of the degree of freedom, $\mbox{df} = (2 n_{\script{otu}} - 3) - 2$,
where $n_{\script{pairs}}$ is the number of site pairs, and
$n_{\script{otu}}$ is the number of OTUs.
\newline
$^c$	TP and FP are the numbers of true and false positives, which are
the number of contact site pairs and the number of non-contact site pairs in each category.
Protein structures used to calculate contact residue pairs are listed
in \Table{\ref{tbl: protein_families}}.
Neighboring residue pairs within 5 residues ($|i - j| \leq 5$)       
along a peptide chain are 
excluded in the evaluation of prediction accuracy.
Also both terminal sites are excluded from counting in this table.
\newline
$^d$	PPV stands for a positive predictive value; i.e., $\mbox{PPV} = \mbox{TP} / (\mbox{TP} + \mbox{FP})$.
} 
}
} 
\end{table}

\begin{table}[hb]
\caption{\label{tbl: pc_each}
\BF{
Co-evolution score
($\rho^x_{ij}$) 
based on each characteristic variable.
} 
}

\vspace*{2em}
\TableInLegends{

\begin{tabular}{lrrrrrr}
\hline
Characteristic	& \multicolumn{3}{c}{$\rho^x_{ij} \geq \rho^s_{ij} \geq r_t$$^a$} & \multicolumn{3}{c}{$ \rho^x_{ij} \leq - \rho^s_{ij} \leq - r_t$$^a$} \\
variable	& TP$^b$ & FP$^b$ & PPV$^c$ & TP	& FP	& PPV	\\
		& \multicolumn{6}{c}{over all protein families}		\\
\hline
Substitution	& 687	& 642 	& 0.52	& 	&	&	\\
Volume		& 18	& 20	& 0.47		& 73	& 10	& \textbf{0.88}$^d$	\\
Charge		& 6	& 8	& 0.43		& 134	& 54	& \textbf{0.71}$^d$	\\
Hydrogen bond	& 4	& 11	& 0.27		& 125	& 51	& \textbf{0.71}$^d$	\\
Hydrophobicity	& 23	& 13	& \textbf{0.64}$^d$	& 23	& 16	& \textbf{0.59}$^d$	\\
$\alpha$ propensity & 14 & 20	& 0.41		& 9 	& 10	& 0.47	\\
$\beta$ propensity & 24	& 17	& \textbf{0.59}$^d$	& 30	& 14	& \textbf{0.68}$^d$	\\
Turn propensity & 21	& 18 	& \textbf{0.54}$^d$	& 17	& 15	& \textbf{0.53}$^d$	\\
Aromatic interaction   & 30	& 10	& \textbf{0.75}$^d$	& 16	& 14	& \textbf{0.53}$^d$	\\
Branched side-chain & 26	& 16	& \textbf{0.62}$^d$	& 20	& 8	& \textbf{0.71}$^d$	\\
Cross link	& 23	& 12	& \textbf{0.66}$^d$	& 5	& 9	& 0.36	\\
Ionic side-chain	& 27	& 15	& \textbf{0.64}$^d$	& 14	& 18	& 0.44	\\
			\\
\hline
\end{tabular}

{
\small{
\vspace*{1em}
$^a$	See \Eqs{\ref{eq: definition_of_effective_pc_s}} and \ref{eq: definition_of_effective_pc_x} 
for the definition of $\rho^x_{ij}$.
The $r_t$ is a threshold for a correlation coefficient corresponding to 
the E-value $E_t = 0.001$ (the P-value $P_t = E_t / n_{\script{pairs}}$),
in the Student's t-distribution of the degree of freedom, $\mbox{df} = (2 n_{\script{otu}} - 3) - 2$,
where $n_{\script{pairs}}$ is the number of site pairs, and
$n_{\script{otu}}$ is the number of OTUs.
\newline
$^b$    TP and FP are the numbers of true and false contact residue pairs; 
protein structures used to calculate contact residue pairs are listed
in \Table{\ref{tbl: protein_families}}.
Neighboring residue pairs within 5 residues ($|i - j| \leq 5$)       
along a peptide chain are excluded in the evaluation of prediction accuracy.
Also both terminal sites are excluded from counting in this table.
\newline
$^c$    PPV stands for a positive predictive value; i.e., $\mbox{PPV} = \mbox{TP} / (\mbox{TP} + \mbox{FP})$.
\newline
$^d$	These PPVs are larger than the PPV for concurrent substitutions, i.e., $0.52$ for $\rho^s$.
}
}
} 
\end{table}

\begin{table}[hb]
\caption{\label{tbl: prediction_accuracy}
\BF{
Accuracy of contact prediction based on the overall co-evolution score ($\rho_{ij}$).
} 
}

\vspace*{2em}
\TableInLegends{

\newcommand{\B}[1]{\textbf{#1}}

\begin{tabular}{lrrrrrrrrrrrrrrr}
\hline
Pfam ID 	&\#contacts	& TP $+$ FP$^b$ & \multicolumn{2}{c}{PPV$^c$}	& \multicolumn{2}{c}{MDPNT$^d$} & \multicolumn{2}{c}{MDTNP$^e$}
				 		\\ 	
		&\hspace*{2em}/\#sites$^a$	&	    & DI $^f$ & $\rho_{ij}$ & DI $^f$ & $\rho_{ij}$ & DI $^f$ & $\rho_{ij}$ 
						\\
\hline
Trans\_reg\_C	&111/76	& 27  	& 0.556	&\B{0.667}& 1.30 &\B{0.94}& 4.20 &\B{3.28}	\\
		&1.5	& 37	& 0.432	&\B{0.622}& 1.72 &\B{1.16}& 3.64 &\B{2.82}	\\
CH		&172/101& 43 	&\B{0.488}&0.465  &\B{2.23}& 2.55 & 4.59 &\B{4.37}	\\
		&1.7	& 57	& 0.439	&\B{0.491}&\B{2.12}& 2.44 & 3.70 &\B{3.30}	\\
7tm\_1		&372/248& 93	& 0.194	&\B{0.344}& 7.43 &\B{5.31}& 12.68 &\B{7.71}	\\
		&1.5	& 124	& 0.169	&\B{0.306}& 7.30 &\B{5.33}& 12.18 &\B{6.40}	\\
\hline
SH3\_1		&89/48	& 22 	& 0.636	&\B{0.682}& 0.83 &\B{0.51}&\B{1.69}& 2.34	\\
		&1.9	& 29	& 0.552	&\B{0.655}& 1.15 &\B{0.62}& 1.56 &\B{1.51}	\\
Cadherin	&220/91	& 55	& 0.818	&\B{0.836}& 0.59 &\B{0.25}& 1.98 & 1.98		\\
		&2.4	& 73	& 0.753	&\B{0.767}& 0.64 &\B{0.45}& 1.60 & 1.60		\\
Trypsin		&636/212& 159 	& 0.591	&\B{0.673}& 1.75 &\B{1.20}& 3.26 &\B{3.10}	\\
		&3.0	& 212	& 0.533	&\B{0.613}& 2.26 &\B{1.65}& 2.83 &\B{1.94}	\\
\hline
Kunitz\_BPTI	&111/53	& 27 	& 0.444	&\B{0.593}& 1.40 &\B{1.18}& 2.31 &\B{2.08}	\\
		&2.1	& 37	&\B{0.541}& 0.486 &\B{1.13} & 1.46&\B{1.86} & 1.94	\\
KH\_1		&90/57	& 22 	& 0.500	&\B{0.773}& 0.99 &\B{0.51}&\B{2.41} & 3.29	\\
		&1.6	& 30	& 0.533	&\B{0.700}& 1.07 &\B{0.56}&\B{2.16} & 3.05	\\
RRM\_1		&133/70	& 33 	& 0.758	&\B{0.818}&\B{0.52} & 0.55& 2.86 &\B{2.36}	\\
		&1.9	& 44	& 0.705	&\B{0.795}& 0.83 &\B{0.49}& 2.49 &\B{1.84}	\\
FKBP\_C		&200/92	& 50 	& 0.760	&\B{0.840}&\B{0.53}& 0.69 & 1.97 &\B{1.85}	\\
		&2.2	& 66	& 0.697	&\B{0.727}& 0.94 &\B{0.85}& 1.66 &\B{1.51}	\\
Lectin\_C	&246/103& 61	&\B{0.770}& 0.705 &\B{0.80} & 0.94& 2.93 &\B{2.67}	\\
		&2.4	& 82	&\B{0.671}& 0.646 & 1.19 &\B{1.17}& 2.54 &\B{2.32}	\\
\hline
Thioredoxin	&188/99	& 47 	& 0.532	&\B{0.638}& 0.98 &\B{0.85}& 3.43 &\B{2.33}	\\
		&1.9	& 62	& 0.565	&\B{0.645}& 0.94 &\B{0.91}& 3.16 &\B{1.86}	\\
Response\_reg	&202/110& 50	& 0.660	&\B{0.680}&\B{0.86} & 0.88& 3.39 &\B{3.06}	\\
		&1.8	& 67	& 0.642	&\B{0.687}& 1.01 &\B{0.92}& 2.54 &\B{2.29}	\\
RNase\_H	&273/128& 68 	&\B{0.559}& 0.471 &\B{1.51}& 1.53&\B{3.61}& 5.44	\\
		&2.1	& 91	&\B{0.549}& 0.407 &\B{1.55}& 2.19& 3.27	&\B{3.07}	\\
Ras		&335/159& 83	& 0.699	& 0.699	  &\B{0.94}& 1.05&\B{2.98}& 3.68	\\
		&2.1	& 111	& 0.631	&\B{0.685}&\B{1.12}& 1.45&\B{2.40}& 2.51	\\
\hline
\end{tabular}

{
\small{
\vspace*{1em}
$^a$	
Protein structures used to calculate contact residue pairs are listed 
in \Table{\ref{tbl: protein_families}}.
Neighboring residue pairs within 5 residues ($|i - j| \leq 5$)
along a peptide chain are not counted as contacts in the evaluation of prediction accuracy.                
\newline
$^b$	TP and FP are the numbers of true and false positives;
only predictions for $\mbox{TP} + \mbox{FP} = \mbox{\#contacts} / 4$ and $\mbox{\#contacts} / 3$
are listed.
\newline
$^c$	PPV stands for a positive predictive value; i.e., $\mbox{PPV} = \mbox{TP} / (\mbox{TP} + \mbox{FP})$.
Better values are typed in a bold font.
\newline
$^d$	MDPNT stands for the mean Euclidean distance from predicted site pairs to the nearest true contact in
the 2-dimensional sequence-position space\CITE{MCSHPZS:11}.
Better values are typed in a bold font.
\newline
$^e$	MDTNP stands for 
the mean Euclidean distance from every true contact to the nearest predicted site pair
in the 2-dimensional sequence-position space\CITE{MCSHPZS:11}.
Better values are typed in a bold font.
\newline
$^f$	DI means the prediction based on the direct information (DI) score published 
in \CITE{MCSHPZS:11}; 
a filtering based on a secondary structure prediction is not applied but
only a conservation filter\CITE{MCSHPZS:11} is.
}
}
} 
\end{table}

} 

\clearpage

\FULLorSHORT{


}{	

\section*{Supporting Information}

\renewcommand{\FigureInLegends}[1]{#1}

\renewcommand{\TextFig}[1]{}
\renewcommand{\SupFig}[1]{#1}
\setcounter{figure}{0}
\renewcommand{\thefigure}{S\arabic{figure}}

\SupFig{

\begin{figure*}[ht]
\vspace*{1em}
\caption{
\label{fig: cmap_for_all_proteins}
\BF{Co-evolving site pairs versus DI residue pairs.}
Residue pairs whose minimum atomic distances are shorter than 5 \AA $\;$
in a protein structure and co-evolving site pairs predicted are shown
by gray filled-squares and by red or indigo filled-circles in the
lower-left half of each figure, respectively.
For comparison, such residue-residue proximities 
and predicted contact residue pairs with high DI scores in \CITE{MCSHPZS:11}
are shown by gray filled-squares and by red or indigo filled-circles 
in the upper-right half of each figure, respectively; 
only the conservation filter is applied but the filters based on
a secondary structure prediction and for cysteine pairs are not applied to the DI scores.
Red and indigo filled-circles correspond to true and false contact residue pairs, respectively.
Residue pairs separated by five or fewer positions in a sequence may be shown with the
gray filled-squares but are excluded in both the predictions.    
The total numbers of co-evolving site pairs and DI residue pairs plotted for each protein
are both equal to one third of true contacts ($\mbox{TP} + \mbox{FP} = \mbox{\#contacts} / 3$).
The PPVs of both the methods for each protein are listed in \Table{\ref{tbl: prediction_accuracy}}.
}
\end{figure*}

\FigureInLegends{\newpage}

\FigureInLegends{

\centerline{
\includegraphics*[width=80mm,angle=0]{FIGS/Trans_reg_C.cmap.eps}
\hspace*{5mm}
\includegraphics*[width=83mm,angle=0]{FIGS/CH.cmap.eps}
}
\vspace*{1em}

\centerline{
\includegraphics*[width=83mm,angle=0]{FIGS/7tm_1.cmap.eps}
\hspace*{5mm}
\hspace*{80mm}
}
} 
\vspace*{1em}

\FigureInLegends{\newpage}

\FigureInLegends{

\centerline{
\includegraphics*[width=80mm,angle=0]{FIGS/SH3_1.cmap.eps}
\hspace*{5mm}
\includegraphics*[width=83mm,angle=0]{FIGS/Cadherin.cmap.eps}
}
\vspace*{1em}

\centerline{
\includegraphics*[width=83mm,angle=0]{FIGS/Trypsin.cmap.eps}
\hspace*{5mm}
\hspace*{80mm}
}
} 
\vspace*{1em}

\FigureInLegends{\newpage}

\FigureInLegends{
\centerline{
\includegraphics*[width=80mm,angle=0]{FIGS/Kunitz_BPTI.cmap.eps}
\hspace*{5mm}
\includegraphics*[width=80mm,angle=0]{FIGS/KH_1.cmap.eps}
}
\vspace*{1em}

\centerline{
\includegraphics*[width=80mm,angle=0]{FIGS/RRM_1.cmap.eps}
\hspace*{5mm}
\includegraphics*[width=80mm,angle=0]{FIGS/FKBP_C.cmap.eps}
}
\vspace*{1em}

\centerline{
\includegraphics*[width=83mm,angle=0]{FIGS/Lectin_C.cmap.eps}
\hspace*{5mm}
\hspace*{80mm}
}
} 
\vspace*{1em}

\FigureInLegends{\newpage}

\FigureInLegends{

\centerline{
\includegraphics*[width=83mm,angle=0]{FIGS/Thioredoxin.cmap.eps}
\hspace*{5mm}
\includegraphics*[width=83mm,angle=0]{FIGS/Response_reg.cmap.eps}
}
\vspace*{1em}

\centerline{
\includegraphics*[width=83mm,angle=0]{FIGS/RNase_H.cmap.eps}
\hspace*{5mm}
\includegraphics*[width=83mm,angle=0]{FIGS/Ras.cmap.eps}
}
} 
\vspace*{1em}

\FigureInLegends{\newpage}

\begin{figure*}[ht]
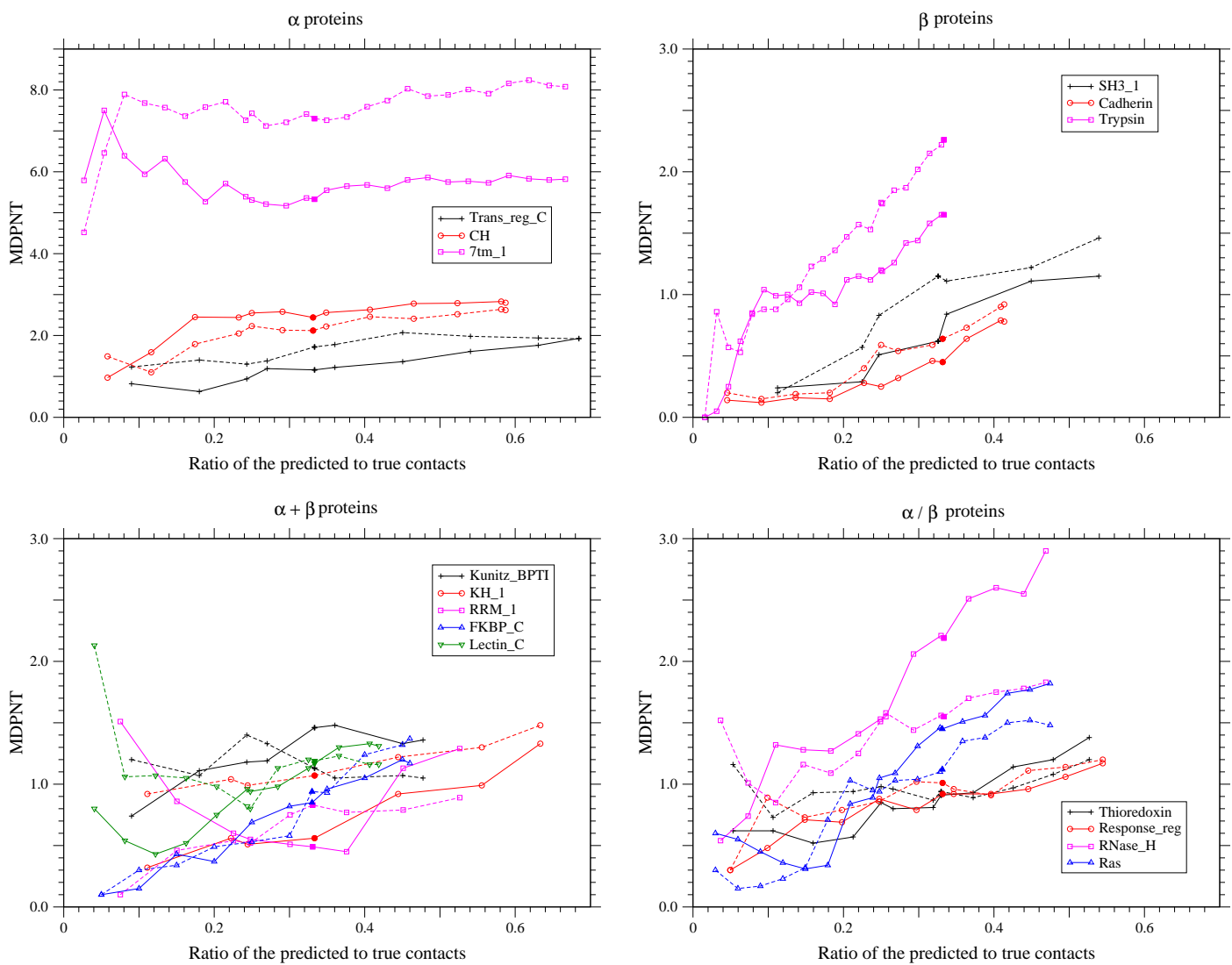

\FigureInLegends{

\centerline{
\includegraphics*[width=90mm,angle=0]{FIGS/MDPNT_vs_nc_over_tnc.a.eps}
\hspace*{3mm}
\includegraphics*[width=90mm,angle=0]{FIGS/MDPNT_vs_nc_over_tnc.b.eps}
}
\vspace*{1em}

\centerline{
\includegraphics*[width=90mm,angle=0]{FIGS/MDPNT_vs_nc_over_tnc.a+b.eps}
\hspace*{3mm}
\includegraphics*[width=90mm,angle=0]{FIGS/MDPNT_vs_nc_over_tnc.a-b.eps}
}
} 
\vspace*{1em}
\caption{
\label{fig: MDPNT_vs_nc_over_tnc}
\BF{Dependence of MDPNT on the number of predicted contacts.}
The dependences of 
the mean Euclidean distance from predicted site pairs to the nearest true contact in
the 2-dimensional sequence-position space
on the total number of predicted contacts
are shown for each protein fold of $\alpha$, $\beta$, $\alpha + \beta$, and $\alpha / \beta$.
The solid and dotted lines show the MDPNTs of the present method and
the method based on the DI score\CITE{MCSHPZS:11}, respectively.
The total number of predicted contacts is shown in the scale of
the ratio of the number of predicted contacts to the number of true contacts.
The total number of predicted site pairs takes every 10
from 10 to a sequence length; also
MDPNTs for the numbers of predicted contacts equal to one fourth or one third
of true contacts are plotted.
The filled marks indicate the points corresponding to the number of predicted site pairs
equal to one third of the number of true contacts.
The number of sequences used here for each protein family is one listed in 
\Table{\ref{tbl: protein_families}}.
}
\end{figure*}

\FigureInLegends{\newpage}

\begin{figure*}[ht]
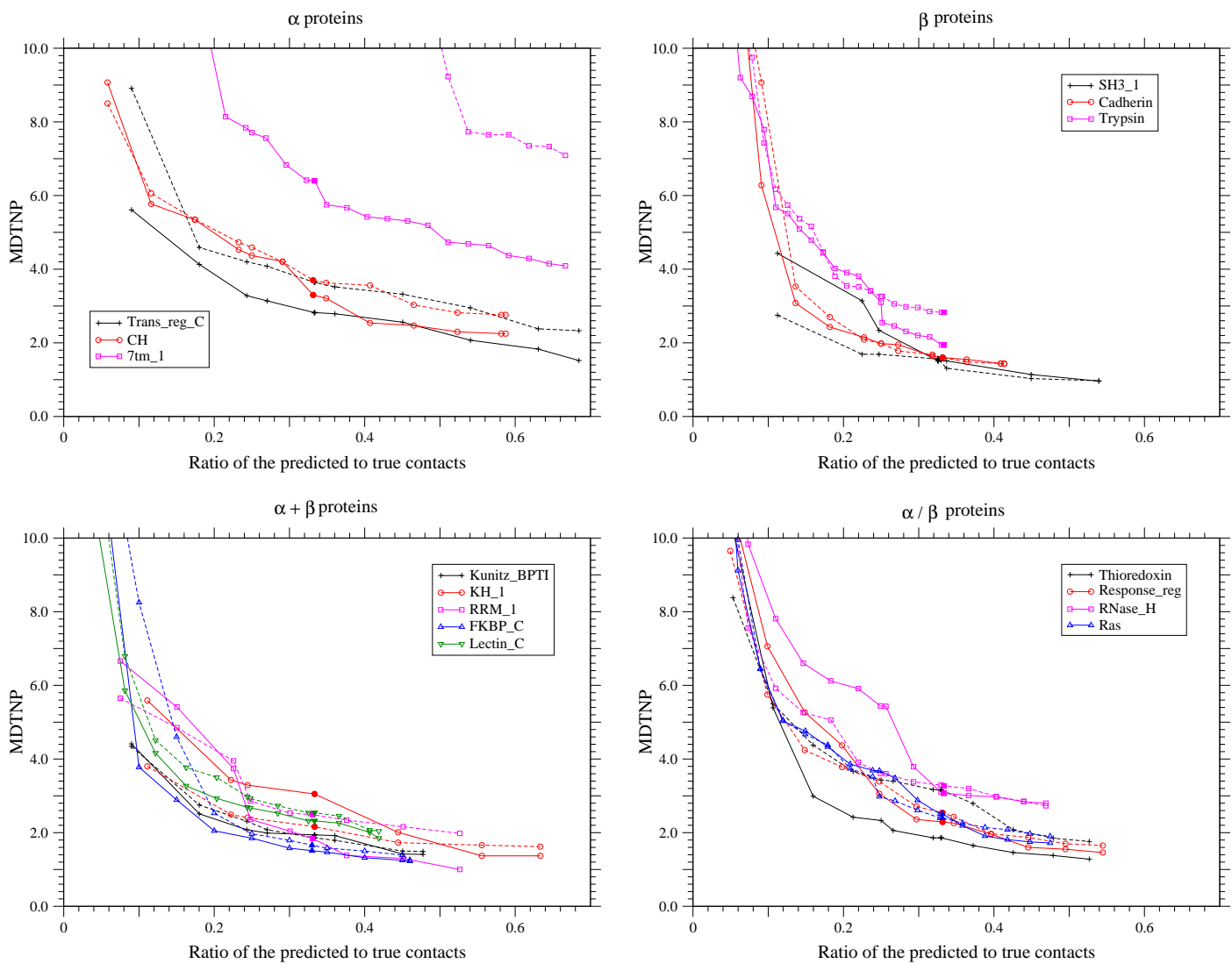

\FigureInLegends{

\centerline{
\includegraphics*[width=90mm,angle=0]{FIGS/MDTNP_vs_nc_over_tnc.a.eps}
\hspace*{3mm}
\includegraphics*[width=90mm,angle=0]{FIGS/MDTNP_vs_nc_over_tnc.b.eps}
}
\vspace*{1em}

\centerline{
\includegraphics*[width=90mm,angle=0]{FIGS/MDTNP_vs_nc_over_tnc.a+b.eps}
\hspace*{3mm}
\includegraphics*[width=90mm,angle=0]{FIGS/MDTNP_vs_nc_over_tnc.a-b.eps}
}
} 
\vspace*{1em}
\caption{
\label{fig: MDTNP_vs_nc_over_tnc}
\BF{Dependence of MDTNP on the number of predicted contacts.}
The dependences of
the mean Euclidean distance from every true contact to the nearest predicted site pair
in the 2-dimensional sequence-position space
on the total number of predicted contacts
are shown for each protein fold of $\alpha$, $\beta$, $\alpha + \beta$, and $\alpha / \beta$.
The solid and dotted lines show the MDTNPs of the present method and
the method based on the DI score\CITE{MCSHPZS:11}, respectively.
The total number of predicted site pairs is shown in the scale of
the ratio of the number of predicted site pairs to the number of true contacts.
The total number of predicted site pairs takes every 10
from 10 to a sequence length; also
MDTNPs for the numbers of predicted site pairs equal to one fourth or one third
of true contacts are plotted.
The filled marks indicate the points corresponding to the number of predicted contacts 
equal to one third of the number of true contacts.
The number of sequences used here for each protein family is one listed in 
\Table{\ref{tbl: protein_families}}.
}
\end{figure*}

\FigureInLegends{\newpage}

\begin{figure*}[ht]
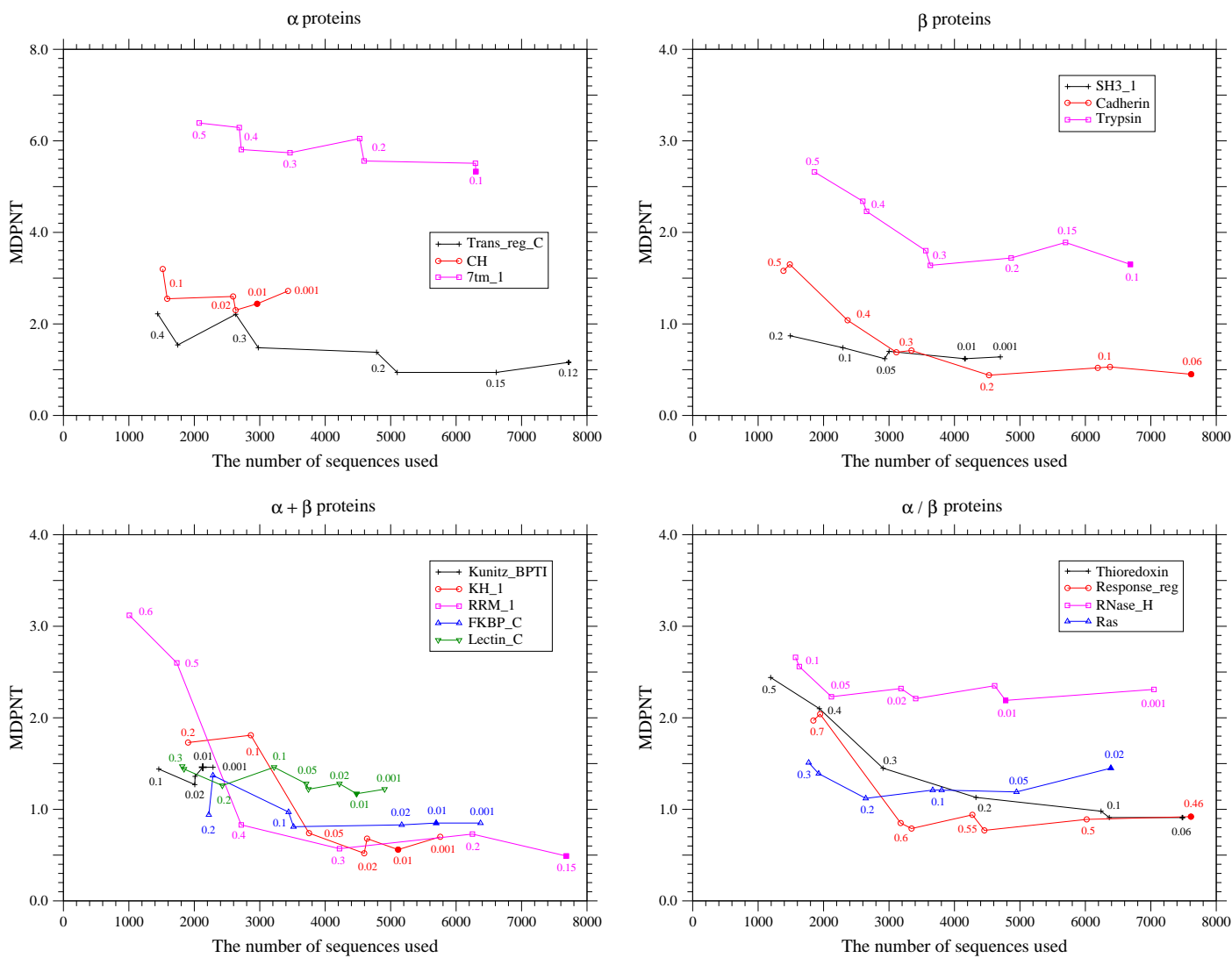

\FigureInLegends{

\centerline{
\includegraphics*[width=90mm,angle=0]{FIGS/MDPNT_vs_n_seqs.a.eps}
\hspace*{3mm}
\includegraphics*[width=90mm,angle=0]{FIGS/MDPNT_vs_n_seqs.b.eps}
}
\vspace*{1em}

\centerline{
\includegraphics*[width=90mm,angle=0]{FIGS/MDPNT_vs_n_seqs.a+b.eps}
\hspace*{3mm}
\includegraphics*[width=90mm,angle=0]{FIGS/MDPNT_vs_n_seqs.a-b.eps}
}
} 
\vspace*{1em}
\caption{
\label{fig: MDPNT_vs_n_seqs}
\BF{Dependence of MDPNT on the number of sequences used.}
The mean Euclidean distance from every predicted site pair to the nearest true contact in
the 2-dimensional sequence-position space
is plotted against
the total number of homologous sequences used
for each prediction.
The filled marks indicate the points corresponding to the number of used sequences
listed for each protein family in \Table{\ref{tbl: protein_families}}.
The values written near each data point indicate the threshold value $T_{bt}$;
OTUs connected to their parent nodes with branches shorter than this
threshold value are removed in the NJ tree of the Pfam full sequences used for each prediction.
Some data points correspond to datasets generated by using the same value of the threshold
but by removing different OTUs.
}
\end{figure*}

\FigureInLegends{\newpage}

\begin{figure*}[ht]
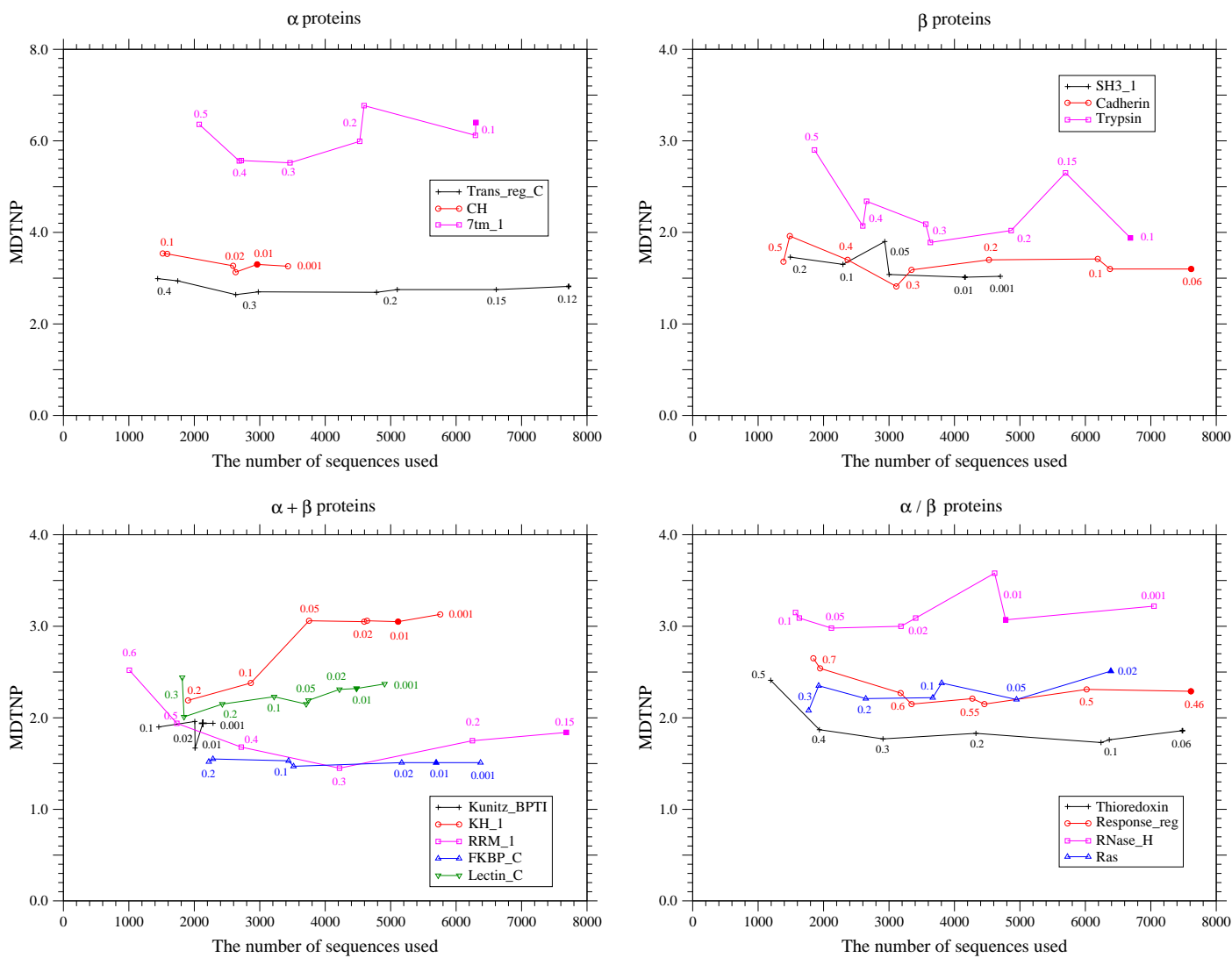

\FigureInLegends{

\centerline{
\includegraphics*[width=90mm,angle=0]{FIGS/MDTNP_vs_n_seqs.a.eps}
\hspace*{3mm}
\includegraphics*[width=90mm,angle=0]{FIGS/MDTNP_vs_n_seqs.b.eps}
}
\vspace*{1em}

\centerline{
\includegraphics*[width=90mm,angle=0]{FIGS/MDTNP_vs_n_seqs.a+b.eps}
\hspace*{3mm}
\includegraphics*[width=90mm,angle=0]{FIGS/MDTNP_vs_n_seqs.a-b.eps}
}
} 
\vspace*{1em}
\caption{
\label{fig: MDTNP_vs_n_seqs}
\BF{Dependence of MDTNP on the number of sequences used.}
The mean Euclidean distance from every true contact to the nearest predicted site pair
in the 2-dimensional sequence-position space
is plotted against
the total number of homologous sequences used
for each prediction.
The filled marks indicate the points corresponding to the number of used sequences
listed for each protein family in \Table{\ref{tbl: protein_families}}.
The values written near each data point indicate the threshold value $T_{bt}$;
OTUs connected to their parent nodes with branches shorter than this
threshold value are removed in the NJ tree of the Pfam full sequences used for each prediction.
Some data points correspond to datasets generated by using the same value of the threshold
but by removing different OTUs.
}
\end{figure*}

} 
\clearpage

\renewcommand{\TableInLegends}[1]{}
\renewcommand{\TextTable}[1]{}
\renewcommand{\SupTable}[1]{#1}
\setcounter{table}{0}
\renewcommand{\thetable}{S\arabic{table}}

\TextTable{

\begin{table}[hb]
\caption{\label{tbl: protein_families}
\BF{
Protein families used.
} 
}

\vspace*{2em}
\TableInLegends{

\begin{tabular}{lrrlllc}
\hline
Pfam ID$^a$     & Seed$^b$ & Full$^c$ &	\multicolumn{2}{c}{Target protein domain} &	Fold type 	& No. sites/Length$^f$	\\
		&	&	& Uniprot ID$^d$	& PDB ID$^e$ &		&	\\
\hline
Trans\_reg\_C	& 362	& 35180	& OMPR\_ECOLI/156-232	& 1ODD-A:156-232	& $\alpha$	& 76/77	\\
CH	& 202	& 5756	& SPTB2\_HUMAN/176-278	& 1BKR-A:5-107	& $\alpha$	& 101/103	\\
7tm\_1	& 64	& 26656	& OPSD\_BOVIN/54-306	& 1GZM-A:54-306	& $\alpha$ (tm)$^g$	& 248/253	\\
SH3\_1	& 61	& 8993	& YES\_HUMAN/97-144	& 2HDA-A:97-144	& $\beta$	& 48/48	\\
Cadherin	& 57	& 18808	& CADH1\_HUMAN/267-366	& 2O72-A:113-212 & $\beta$	& 91/100	\\
Trypsin	& 71	& 14720	& TRY2\_RAT/24-239	& 3TGI-E:16-238	& $\beta$	& 212/216	\\
Kunitz\_BPTI	& 151	& 3090	& BPT1\_BOVIN/39-91	& 5PTI-A:4-56	& $\alpha+\beta$	& 53/53	\\
KH\_1	& 399	& 11484	& PCBP1\_HUMAN/281-343	& 1WVN-A:7-69	& $\alpha+\beta$	& 57/63	\\
RRM\_1	& 79	& 31837	& ELAV4\_HUMAN/48-118	& 1G2E-A:41-111	& $\alpha+\beta$	& 70/71	\\
FKBP\_C	& 174	& 11034	& O45418\_CAEEL/26-118	& 1R9H-A:26-118	& $\alpha+\beta$	& 92/93	\\
Lectin\_C	& 44	& 6530	& CD209\_HUMAN/273-379	& 1SL5-A:273-379	& $\alpha+\beta$	& 103/107	\\
Thioredoxin	& 50	& 16281	& THIO\_ALIAC/1-103	& 1RQM-A:1-103	& $\alpha/\beta$ 	& 99/103	\\
Response\_reg	& 57	& 103232	& CHEY\_ECOLI/8-121	& 1E6K-A:8-121	& $\alpha/\beta$	& 110/114	\\
RNase\_H	& 65	& 13801	& RNH\_ECOLI/2-142	& 1F21-A:3-142 & $\alpha/\beta$	& 128/140	\\
Ras	& 61	& 13525	& RASH\_HUMAN/5-165	&	5P21-A:5-165	& $\alpha/\beta$	& 159/161	\\
\hline
\end{tabular}

{
\small{
\vspace*{1em}
\renewcommand{\baselinestretch}{1.0}
$^a$  Pfam release 26.0 (November 2011) was used.
\newline
$^b$  The number of sequences included in the seed alignment of the Pfam.
\newline
$^c$  The number of sequences included in the full alignment of the Pfam.
\newline
$^d$  Target protein domain in the Pfam family.
\newline
$^e$  A protein structure corresponding to the target protein domain.
\newline
$^f$  Unreliable site positions that are represented by the lower case of characters 
in alignments were excluded in the evaluation of prediction accuracy.
\newline
$^g$ Transmembrane $\alpha$.
}
}

} 
\end{table}

\begin{table}[hb]
\caption{\label{tbl: c_vs_pc}
\BF{
Correlation ($r_{\Delta^s_i \Delta^s_j}$) versus 
partial correlation ($r_{\prod_{\perp}\Delta^s_i \prod_{\perp}\Delta^s_j}$) 
coefficients of concurrent substitutions between sites.
} 
}

\vspace*{2em}
\TableInLegends{

\begin{tabular}{lrrrrrrrrrr}
\hline
Pfam ID & $T_{bt}$$^a$	& \#seqs$^a$ & \multicolumn{2}{c}{$C^s_{ij} \geq r_t$$^b$} & \multicolumn{2}{c}{$r_t > C^s_{ij} > 0$}  
				& \multicolumn{2}{c}{$0 > C^s_{ij} > - r_t$} & \multicolumn{2}{c}{$-r_t \geq C^s_{ij}$}	\\
	 & 		& 	 & TP:FP$^c$ & PPV$^d$ & TP:FP & PPV & TP:FP & PPV & TP:FP & PPV	\\
\hline
Trans\_reg\_C	& 0.12	& 7720	& 102:2282 &0.04& 1:30	& 0.03	& 0:0	& --	& 0:0	& --	\\
CH		& 0.01	& 2960	& 167:4226 &0.04& 2:73	&0.03	& 0:2	& 0.0	& 0:0	& --	\\
7tm\_1		& 0.1	& 6302	& 358:28576&0.01& 0:0	& --	& 0:0	& --	& 0:0	& --	\\
SH3\_1		& 0.01	& 4160	& 74:674 &0.10	& 7:60	& 0.10	& 0:5	& 0.0	& 0:0	& --	\\
Cadherin	& 0.06	& 7617	& 214:3333&0.06& 1:46	&0.02	& 0:7	& 0.0	& 0:0	& --	\\
Trypsin		& 0.1	& 6688	& 617:20312&0.03& 0:0	& --	& 0:0	& --	& 0:0	& --	\\
Kunitz\_BPTI	& 0.01	& 2130	& 86:799 &0.10	& 11:48	&0.19	& 0:2	& 0.0	& 0:0	& --	\\
KH\_1		& 0.01	& 5114	& 78:1116&0.07	& 1:41	&0.02	& 0:4	& 0.0	& 0:0	& --	\\
RRM\_1		& 0.15	& 7684	& 119:1839&0.06	& 0:0	& --	& 0:0	& --	& 0:0	& --	\\
FKBP\_C		& 0.01	& 5695	& 199:3445&0.05	& 0:10	& 0.0	& 0:1	& 0.0	& 0:0	& --	\\
Lectin\_C	& 0.01	& 4479	& 234:4319&0.05	& 1:19	&0.05	& 0:0	& --	& 0:0	& --	\\
Thioredoxin	& 0.06	& 7483	& 188:4180&0.04	& 0:3	&0.0	& 0:0	& --	& 0:0	& --	\\
Response\_reg	& 0.46	& 7613	& 202:5266&0.04	& 0:1	& 0.0	& 0:0	& --	& 0:0	& --	\\
RNase\_H	& 0.01	& 4782	& 271:7152&0.04	& 0:5	& 0.0 	& 0:0	& --	& 0:0	& --	\\
Ras		& 0.02	& 6390	& 329:11304&0.03& 0:0	& --	& 0:0	& --	& 0:0	& --	\\
\hline
\hline
Pfam ID & \multicolumn{2}{l}{\#contacts} & \multicolumn{2}{c}{$\mathcal{C}^s_{ij} \geq r_t$$^b$} & \multicolumn{2}{c}{$r_t > \mathcal{C}^s_{ij} > 0$}  
				& \multicolumn{2}{c}{$0 > \mathcal{C}^s_{ij} > - r_t$} & \multicolumn{2}{c}{$-r_t \geq  \mathcal{C}^s_{ij}$}	\\
 & \multicolumn{2}{l}{\hspace*{1ex}}/\#sites$^c$  & TP:FP$^c$ & PPV$^d$ & TP:FP & PPV & TP:FP & PPV & TP:FP & PPV	\\
\hline
Trans\_reg\_C	& 103/75& 1.4 	& 32:57	&0.36	& 59:1584&0.04	& 12:669&0.02	& 0:2	& 0.0	\\
CH		& 169/100&  1.7	& 16:17	&0.48	& 125:2454&0.05	& 28:1828&0.02	& 0:2	& 0.0	\\
7tm\_1		& 366/247& 1.5	& 36:84	&0.30	&263:15695&0.02	&59:12787&0.005	& 0:10	& 0.0	\\
SH3\_1		& 81/46	& 1.8	& 24:17	&0.59	& 46:516&0.08	&11:206	&0.05	& 0:0	& --	\\
Cadherin	& 215/90& 2.4	& 40:8	& 0.83	&132:1519&0.08	&42:1857& 0.02	& 1:2	& 0.33	\\
Trypsin		& 617/210& 2.9	& 115:75&0.61 	& 383:11331 &0.03& 119:8899 &0.01& 0:7	& 0.0	\\
Kunitz\_BPTI	& 105/51& 2.1	& 16:12	&0.57	& 55:575&0.09	&26:262	& 0.09	& 0:0	& --	\\
KH\_1		& 79/55	& 1.4	& 19:15	&0.56	& 50:707&0.07	& 10:438& 0.02	& 0:1	& 0.0	\\
RRM\_1		& 119/68& 1.8	& 45:36	&0.56	& 63:1257&0.05	& 11:546&0.02	& 0:0	& --	\\
FKBP\_C		& 199/91& 2.2	& 66:51	&0.56	& 103:2114&0.05	& 30:1288&0.02	& 0:3	& 0.0	\\
Lectin\_C	& 243/102& 2.4	& 36:13	& 0.73	& 160:2401&0.06	& 39:1923& 0.02	& 0:1	& 0.0	\\
Thioredoxin	& 188/99& 1.9	& 53:61	&0.46	& 109:2677&0.04	& 26:1442&0.02	& 0:3	& 0.0	\\
Response\_reg	& 202/110& 1.8	& 72:87 &0.45	& 101:3182&0.03	& 28:1988&0.01	& 1:10	&0.09	\\
RNase\_H	& 271/127& 2.1	& 37:56	&0.40	& 161:3700&0.04	& 72:3387&0.02	& 1:14	&0.07	\\
Ras		& 329/158& 2.1	& 81:55	&0.60	& 203:6472&0.03	& 44:4768&0.01	& 1:9	&0.10	\\
\hline
\end{tabular}

{
\small{
\vspace*{1em}
$^a$	OTUs connected to their parent nodes with branches shorter than 
the threshold value $T_{bt}$ are removed from each Pfam full alignment, 
and the number of remaining OTUs is listed.
\newline
$^b$	The $r_t$ is a threshold for a correlation coefficient corresponding to
the E-value $E_t = 0.001$ (the P-value $P_t = E_t / n_{\script{pairs}}$)
in the Student's t-distribution of the degree of freedom, $\mbox{df} = (2 n_{\script{otu}} - 3) - 2$,
where $n_{\script{pairs}}$ is the number of site pairs, and
$n_{\script{otu}}$ is the number of OTUs.
\newline
$^c$	TP and FP are the numbers of true and false positives, which are
the number of contact site pairs and the number of non-contact site pairs in each category.
Protein structures used to calculate contact residue pairs are listed
in \Table{\ref{tbl: protein_families}}.
Neighboring residue pairs within 5 residues ($|i - j| \leq 5$)       
along a peptide chain are 
excluded in the evaluation of prediction accuracy.
Also both terminal sites are excluded from counting in this table.
\newline
$^d$	PPV stands for a positive predictive value; i.e., $\mbox{PPV} = \mbox{TP} / (\mbox{TP} + \mbox{FP})$.
} 
}
} 
\end{table}

\begin{table}[hb]
\caption{\label{tbl: pc_each}
\BF{
Co-evolution score
($\rho^x_{ij}$) 
based on each characteristic variable.
} 
}

\vspace*{2em}
\TableInLegends{

\begin{tabular}{lrrrrrr}
\hline
Characteristic	& \multicolumn{3}{c}{$\rho^x_{ij} \geq \rho^s_{ij} \geq r_t$$^a$} & \multicolumn{3}{c}{$ \rho^x_{ij} \leq - \rho^s_{ij} \leq - r_t$$^a$} \\
variable	& TP$^b$ & FP$^b$ & PPV$^c$ & TP	& FP	& PPV	\\
		& \multicolumn{6}{c}{over all protein families}		\\
\hline
Substitution	& 687	& 642 	& 0.52	& 	&	&	\\
Volume		& 18	& 20	& 0.47		& 73	& 10	& \textbf{0.88}$^d$	\\
Charge		& 6	& 8	& 0.43		& 134	& 54	& \textbf{0.71}$^d$	\\
Hydrogen bond	& 4	& 11	& 0.27		& 125	& 51	& \textbf{0.71}$^d$	\\
Hydrophobicity	& 23	& 13	& \textbf{0.64}$^d$	& 23	& 16	& \textbf{0.59}$^d$	\\
$\alpha$ propensity & 14 & 20	& 0.41		& 9 	& 10	& 0.47	\\
$\beta$ propensity & 24	& 17	& \textbf{0.59}$^d$	& 30	& 14	& \textbf{0.68}$^d$	\\
Turn propensity & 21	& 18 	& \textbf{0.54}$^d$	& 17	& 15	& \textbf{0.53}$^d$	\\
Aromatic interaction   & 30	& 10	& \textbf{0.75}$^d$	& 16	& 14	& \textbf{0.53}$^d$	\\
Branched side-chain & 26	& 16	& \textbf{0.62}$^d$	& 20	& 8	& \textbf{0.71}$^d$	\\
Cross link	& 23	& 12	& \textbf{0.66}$^d$	& 5	& 9	& 0.36	\\
Ionic side-chain	& 27	& 15	& \textbf{0.64}$^d$	& 14	& 18	& 0.44	\\
			\\
\hline
\end{tabular}

{
\small{
\vspace*{1em}
$^a$	See \Eqs{\ref{eq: definition_of_effective_pc_s}} and \ref{eq: definition_of_effective_pc_x} 
for the definition of $\rho^x_{ij}$.
The $r_t$ is a threshold for a correlation coefficient corresponding to 
the E-value $E_t = 0.001$ (the P-value $P_t = E_t / n_{\script{pairs}}$),
in the Student's t-distribution of the degree of freedom, $\mbox{df} = (2 n_{\script{otu}} - 3) - 2$,
where $n_{\script{pairs}}$ is the number of site pairs, and
$n_{\script{otu}}$ is the number of OTUs.
\newline
$^b$    TP and FP are the numbers of true and false contact residue pairs; 
protein structures used to calculate contact residue pairs are listed
in \Table{\ref{tbl: protein_families}}.
Neighboring residue pairs within 5 residues ($|i - j| \leq 5$)       
along a peptide chain are excluded in the evaluation of prediction accuracy.
Also both terminal sites are excluded from counting in this table.
\newline
$^c$    PPV stands for a positive predictive value; i.e., $\mbox{PPV} = \mbox{TP} / (\mbox{TP} + \mbox{FP})$.
\newline
$^d$	These PPVs are larger than the PPV for concurrent substitutions, i.e., $0.52$ for $\rho^s$.
}
}
} 
\end{table}

\begin{table}[hb]
\caption{\label{tbl: prediction_accuracy}
\BF{
Accuracy of contact prediction based on the overall co-evolution score ($\rho_{ij}$).
} 
}

\vspace*{2em}
\TableInLegends{

\newcommand{\B}[1]{\textbf{#1}}

\begin{tabular}{lrrrrrrrrrrrrrrr}
\hline
Pfam ID 	&\#contacts	& TP $+$ FP$^b$ & \multicolumn{2}{c}{PPV$^c$}	& \multicolumn{2}{c}{MDPNT$^d$} & \multicolumn{2}{c}{MDTNP$^e$}
				 		\\ 	
		&\hspace*{2em}/\#sites$^a$	&	    & DI $^f$ & $\rho_{ij}$ & DI $^f$ & $\rho_{ij}$ & DI $^f$ & $\rho_{ij}$ 
						\\
\hline
Trans\_reg\_C	&111/76	& 27  	& 0.556	&\B{0.667}& 1.30 &\B{0.94}& 4.20 &\B{3.28}	\\
		&1.5	& 37	& 0.432	&\B{0.622}& 1.72 &\B{1.16}& 3.64 &\B{2.82}	\\
CH		&172/101& 43 	&\B{0.488}&0.465  &\B{2.23}& 2.55 & 4.59 &\B{4.37}	\\
		&1.7	& 57	& 0.439	&\B{0.491}&\B{2.12}& 2.44 & 3.70 &\B{3.30}	\\
7tm\_1		&372/248& 93	& 0.194	&\B{0.344}& 7.43 &\B{5.31}& 12.68 &\B{7.71}	\\
		&1.5	& 124	& 0.169	&\B{0.306}& 7.30 &\B{5.33}& 12.18 &\B{6.40}	\\
\hline
SH3\_1		&89/48	& 22 	& 0.636	&\B{0.682}& 0.83 &\B{0.51}&\B{1.69}& 2.34	\\
		&1.9	& 29	& 0.552	&\B{0.655}& 1.15 &\B{0.62}& 1.56 &\B{1.51}	\\
Cadherin	&220/91	& 55	& 0.818	&\B{0.836}& 0.59 &\B{0.25}& 1.98 & 1.98		\\
		&2.4	& 73	& 0.753	&\B{0.767}& 0.64 &\B{0.45}& 1.60 & 1.60		\\
Trypsin		&636/212& 159 	& 0.591	&\B{0.673}& 1.75 &\B{1.20}& 3.26 &\B{3.10}	\\
		&3.0	& 212	& 0.533	&\B{0.613}& 2.26 &\B{1.65}& 2.83 &\B{1.94}	\\
\hline
Kunitz\_BPTI	&111/53	& 27 	& 0.444	&\B{0.593}& 1.40 &\B{1.18}& 2.31 &\B{2.08}	\\
		&2.1	& 37	&\B{0.541}& 0.486 &\B{1.13} & 1.46&\B{1.86} & 1.94	\\
KH\_1		&90/57	& 22 	& 0.500	&\B{0.773}& 0.99 &\B{0.51}&\B{2.41} & 3.29	\\
		&1.6	& 30	& 0.533	&\B{0.700}& 1.07 &\B{0.56}&\B{2.16} & 3.05	\\
RRM\_1		&133/70	& 33 	& 0.758	&\B{0.818}&\B{0.52} & 0.55& 2.86 &\B{2.36}	\\
		&1.9	& 44	& 0.705	&\B{0.795}& 0.83 &\B{0.49}& 2.49 &\B{1.84}	\\
FKBP\_C		&200/92	& 50 	& 0.760	&\B{0.840}&\B{0.53}& 0.69 & 1.97 &\B{1.85}	\\
		&2.2	& 66	& 0.697	&\B{0.727}& 0.94 &\B{0.85}& 1.66 &\B{1.51}	\\
Lectin\_C	&246/103& 61	&\B{0.770}& 0.705 &\B{0.80} & 0.94& 2.93 &\B{2.67}	\\
		&2.4	& 82	&\B{0.671}& 0.646 & 1.19 &\B{1.17}& 2.54 &\B{2.32}	\\
\hline
Thioredoxin	&188/99	& 47 	& 0.532	&\B{0.638}& 0.98 &\B{0.85}& 3.43 &\B{2.33}	\\
		&1.9	& 62	& 0.565	&\B{0.645}& 0.94 &\B{0.91}& 3.16 &\B{1.86}	\\
Response\_reg	&202/110& 50	& 0.660	&\B{0.680}&\B{0.86} & 0.88& 3.39 &\B{3.06}	\\
		&1.8	& 67	& 0.642	&\B{0.687}& 1.01 &\B{0.92}& 2.54 &\B{2.29}	\\
RNase\_H	&273/128& 68 	&\B{0.559}& 0.471 &\B{1.51}& 1.53&\B{3.61}& 5.44	\\
		&2.1	& 91	&\B{0.549}& 0.407 &\B{1.55}& 2.19& 3.27	&\B{3.07}	\\
Ras		&335/159& 83	& 0.699	& 0.699	  &\B{0.94}& 1.05&\B{2.98}& 3.68	\\
		&2.1	& 111	& 0.631	&\B{0.685}&\B{1.12}& 1.45&\B{2.40}& 2.51	\\
\hline
\end{tabular}

{
\small{
\vspace*{1em}
$^a$	
Protein structures used to calculate contact residue pairs are listed 
in \Table{\ref{tbl: protein_families}}.
Neighboring residue pairs within 5 residues ($|i - j| \leq 5$)
along a peptide chain are not counted as contacts in the evaluation of prediction accuracy.                
\newline
$^b$	TP and FP are the numbers of true and false positives;
only predictions for $\mbox{TP} + \mbox{FP} = \mbox{\#contacts} / 4$ and $\mbox{\#contacts} / 3$
are listed.
\newline
$^c$	PPV stands for a positive predictive value; i.e., $\mbox{PPV} = \mbox{TP} / (\mbox{TP} + \mbox{FP})$.
Better values are typed in a bold font.
\newline
$^d$	MDPNT stands for the mean Euclidean distance from predicted site pairs to the nearest true contact in
the 2-dimensional sequence-position space\CITE{MCSHPZS:11}.
Better values are typed in a bold font.
\newline
$^e$	MDTNP stands for 
the mean Euclidean distance from every true contact to the nearest predicted site pair
in the 2-dimensional sequence-position space\CITE{MCSHPZS:11}.
Better values are typed in a bold font.
\newline
$^f$	DI means the prediction based on the direct information (DI) score published 
in \CITE{MCSHPZS:11}; 
a filtering based on a secondary structure prediction is not applied but
only a conservation filter\CITE{MCSHPZS:11} is.
}
}
} 
\end{table}

} 


} 

\end{document}